 \documentclass[12pt,preprint]{aastex} % default

\newcommand{\mhz}{\,$\mu{\rm Hz}$}
\newcommand{\milhz}{\,mHz}
\newcommand{\etal}{et~al.}

\newcommand{\ie}{i.e.} 
\newcommand{\eg}{e.g.}

\newcommand{\cf}{{cf.}}

\newcommand{\topp}{\emph{top}}
\newcommand{\midd}{\emph{middle}}
\newcommand{\bott}{\emph{bottom}}
\newcommand{\upp}{\emph{upper}}

\newcommand{\rii}{\emph{right}}
\newcommand{\lee}{\emph{left}}

\slugcomment{To appear in ApJ., XXX}

\shorttitle{Granulation in Procyon}
\shortauthors{Bruntt et al.}

\begin{document}


\title{Evidence for Granulation and Oscillations in Procyon from Photometry with the WIRE satellite}


\author{H. Bruntt}
\affil{Niels Bohr Institute, University of Copenhagen, Juliane Maries Vej 30,
DK-2100 Copenhagen \O, Denmark}
\email{bruntt@phys.au.dk}            % , Derek.Buzasi@usafa.af.mil}

\author{H. Kjeldsen}
\affil{Department of Physics and Astronomy, University of Aarhus, 
Ny Munkegade, Bygn.\ 520, DK-8000 Aarhus C., Denmark}
\email{hans@phys.au.dk}

\author{D. L. Buzasi}
\affil{US Air Force Academy, Department of Physics, CO, USA}
\email{Derek.Buzasi@usafa.af.mil}

\author{T. R. Bedding}
\affil{School of Physics, University of Sydney, Australia}
\email{T.Bedding@physics.usyd.edu.au}

%%% The latter depends on the assumed lifetime of the oscillations. 
%%% HK: this statement is not true since we work in smoothed PDS: 
%%% but it depends on the number of modes you have included... 

\begin{abstract}
We report evidence for the granulation signal in the 
star Procyon~A, based on two photometric time series 
from the star tracker on the WIRE satellite. 
% which are separated in time by one year. 
The power spectra show
evidence of excess power around 1\milhz, consistent with the
detection of p-modes reported from radial velocity measurements. 
We see a significant increase in the noise level 
below 3\milhz, which we interpret as
the granulation signal.
We have made a large set of numerical simulations to constrain 
the amplitude and timescale of the granulation signal and 
the amplitude of the oscillations. We find that the 
timescale for granulation is $\Gamma_{\rm gran} = 750\pm200$\,s, 
the granulation amplitude is $1.8 \pm 0.3$ times solar, 
and the amplitude of the p-modes is $8\pm3$\,ppm.
We found the distribution of peak heights in 
the observed power spectra to be consistent with that expected from
p-mode oscillations.
However, the quality of the data is not sufficient to measure 
the large separation or detect a comb-like structure, as 
seen in the p-modes of the Sun.
Comparison with the recent negative result from the MOST satellite 
reveal that the MOST data must have an additional 
noise source that prevented the detection of oscillations.
\end{abstract}


%% subject header.  Objects should be in the appropriate "individual"
%% headers (e.g. quasars: individual, stars: individual, etc.) with the
%% additional provision that the total number of headers, including each
%% individual object, not exceed six.  The \objectname{} macro, and its
%% alias \object{}, is used to mark each object.  The macro takes the object
%% name as its primary argument.  This name will appear in the paper
%% and serve as the link's anchor in the electronic edition if the name
%% is recognized by the data centers.  The macro also takes an optional
%% argument in parentheses in cases where the data center identification
%% differs from what is to be printed in the paper.



% ===========================================================================
\section{Introduction}
% ===========================================================================

A clear signature of the transportation of 
heat by convection in the outer parts of the Sun is the
convection cells or granules.
The spatially and temporally incoherent signal generated
by the granulation can be measured as background noise in
the integrated light from the disk of the Sun. 
The physical phenomena on the surface give rise to noise in 
different frequency ranges. In order of increasing 
frequency, we find 
activity ($<1$\,\mhz), 
super-granulation ($<10$\,\mhz),
meso-granulation ($<100$ \mhz)
and granulation ($2\,000-10\,000$\,\mhz) \citep{harvey, aigrain}.

There are several reasons
why it is important to characterize granulation on other stars, apart from
the obvious interest in testing models of convection.
Although the solar five-minute oscillations
($\nu \simeq 3\,000$\,\mhz) are detectable above
the granulation noise, it
difficult to measure solar-like oscillations in
stars with more vigorous surface convection.
Also, the background noise at low frequencies ($< 200$\,\mhz) is 
an important limiting factor when searching for the gravity 
modes in the Sun and other stars. 
Finally, the search for Earth-sized planets 
from space will also be affected by the background 
noise from large-scale granulation \citep{aigrain}.

Evidence for granulation noise in two stars ($\alpha$~Cen~A and Procyon~A)
was reported by \citet{KBF99}, based on measuring the equivalent widths of
temperature-sensitive spectral lines.  More recently,
the MOST satellite \citep{walker03} observed the F5IV type 
star Procyon~A in early 2004 for 32 days 
with a duty cycle of 99\% \citep{matthews04}. 
Since this star is more evolved and hotter than 
the Sun, the granulation is expected to be more vigorous 
and this was indeed what was reported by \citet{matthews04}.
The observed noise in the light curve obtained with MOST 
is four to five times higher that what is measured in the Sun 
in the frequency range 1-2 \milhz\ \citep{bedding05} 
and hampers the detection of the p~modes \citep{matthews04}. 
However, \citet{bedding05} have made simulations of the MOST
time series and suggested that the noise signal seen by \citet{matthews04}
is a combination of granulation and 
instrumental effects (\eg\ scattered light). 

In this study we have used two photometric time-series 
of Procyon~A from the star tracker on 
the Wide-field Infrared Explorer satellite (WIRE).
In Section~\ref{sec:obs} we describe the observations and 
data reduction procedure. In Section~\ref{sec:pow} we
analyze the resulting power density spectra. 
In Section~\ref{sec:sim} we compare the observed power spectra
with a grid of simulations in order to estimate the
amplitude and timescale of the granulation and to 
constrain the amplitude %%% and lifetimes 
of the p~mode signal.
In Section~\ref{sec:con} we discuss our results.

% The MOST satellite has a 15 cm aperture compared to 
% the 5.2 cm aperture of WIRE.

% Procyon is of earlier spectral type than the Sun, 
% \ie\ F5IV-V and its estimated physical parameters 
% are $M/M_\odot=1.5$, \teff$=6500$\,K.

% ===========================================================================
\section{Observations and Data Reduction\label{sec:obs}}
% ===========================================================================

%%% Software for plots: procyon_plots2.pro
  \begin{figure}
   \epsscale{.80}
   \plotone{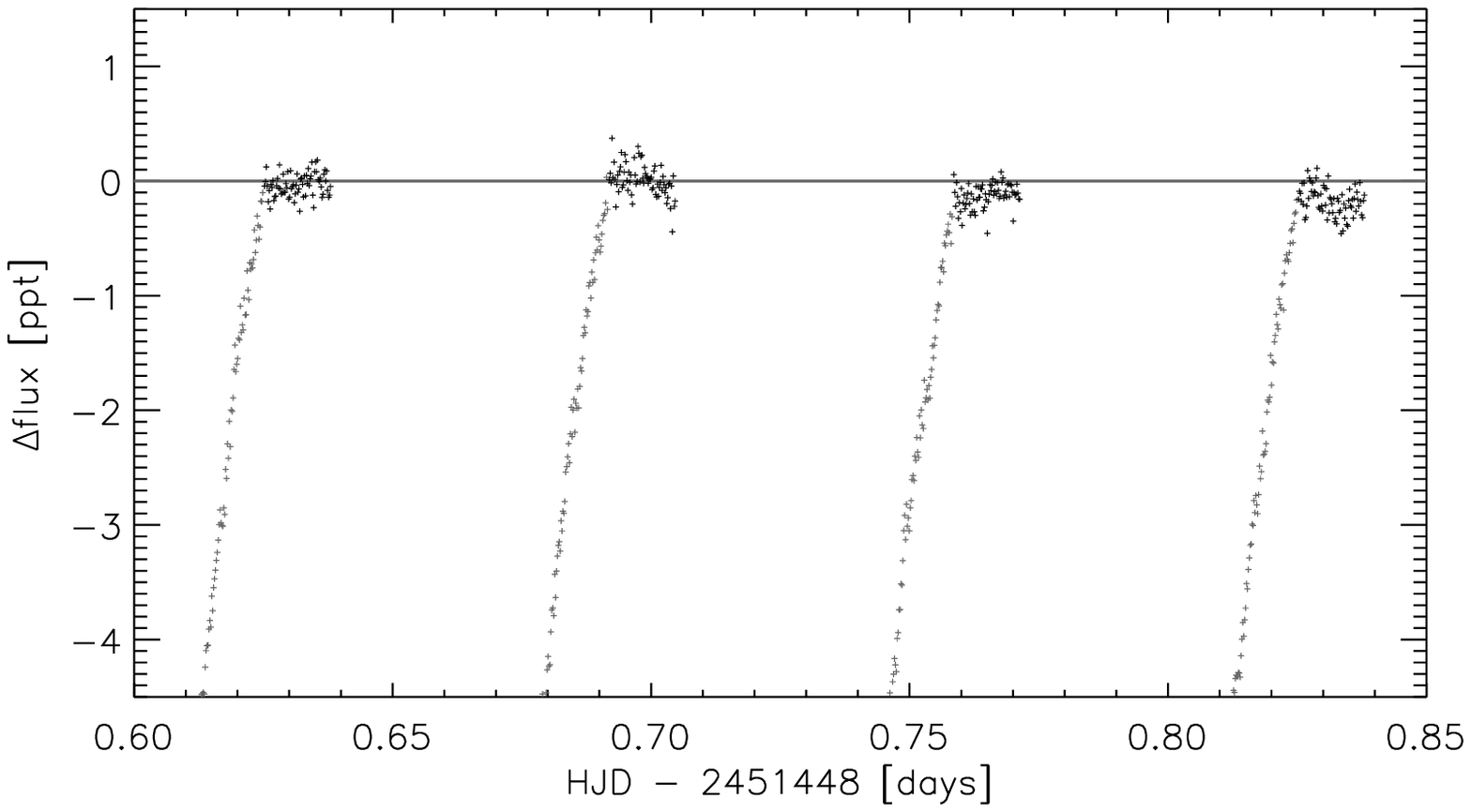} % lc1999_datuse.eps}
   \plotone{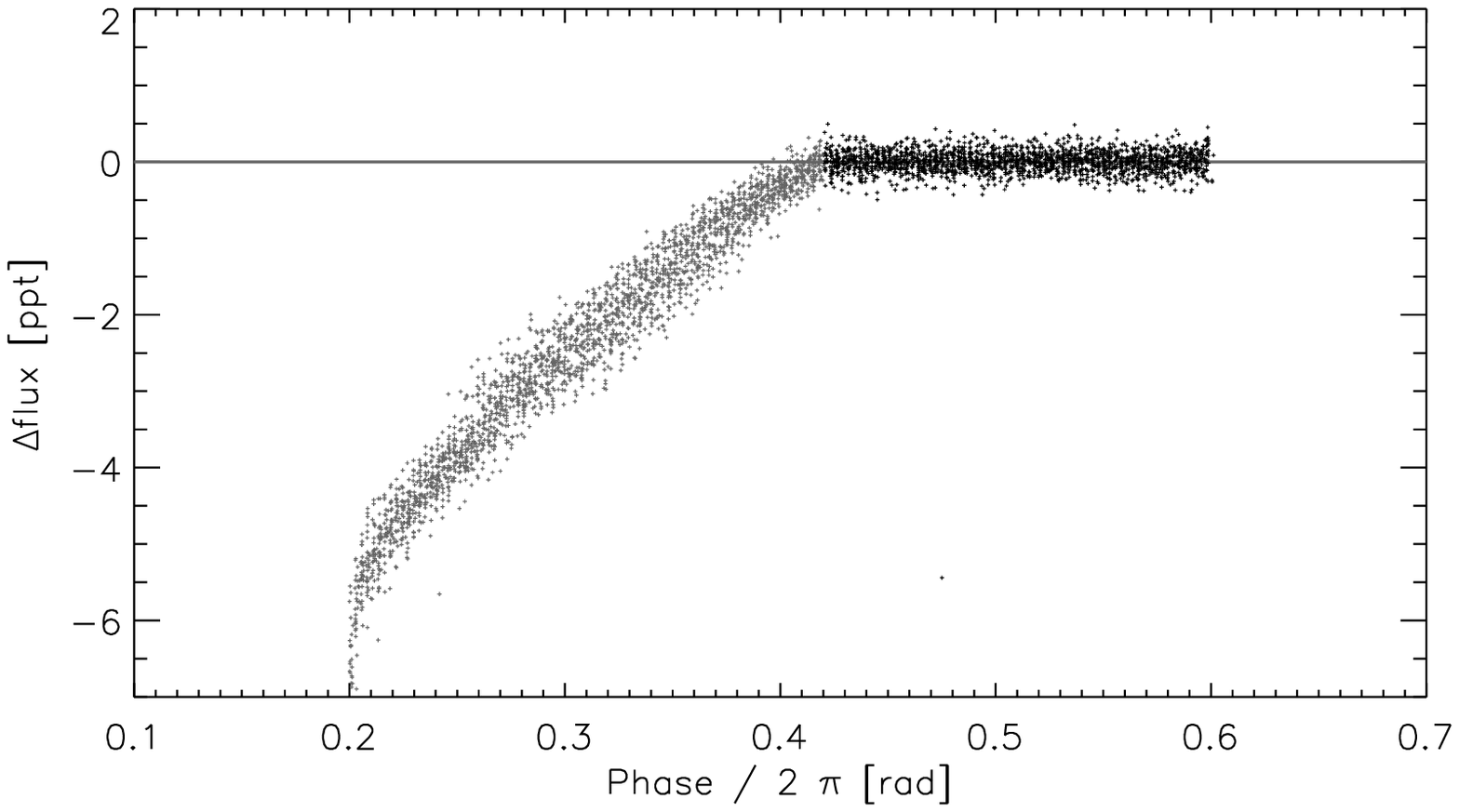} % phase1999.eps}
   \caption{The \emph{top} plot shows part of the raw light curve of 
Procyon from 1999. 
The gray data points at the beginning of each orbit of WIRE
are affected by scattered light. The \emph{bottom} plot shows
the phased light curve (only every 20th data point in plotted). The
gray points mark the points that are affected by scattered light and
which were not used in the analysis.
    \label{fig1}}
   \end{figure}

   \begin{figure}
   \epsscale{.80}
   \plotone{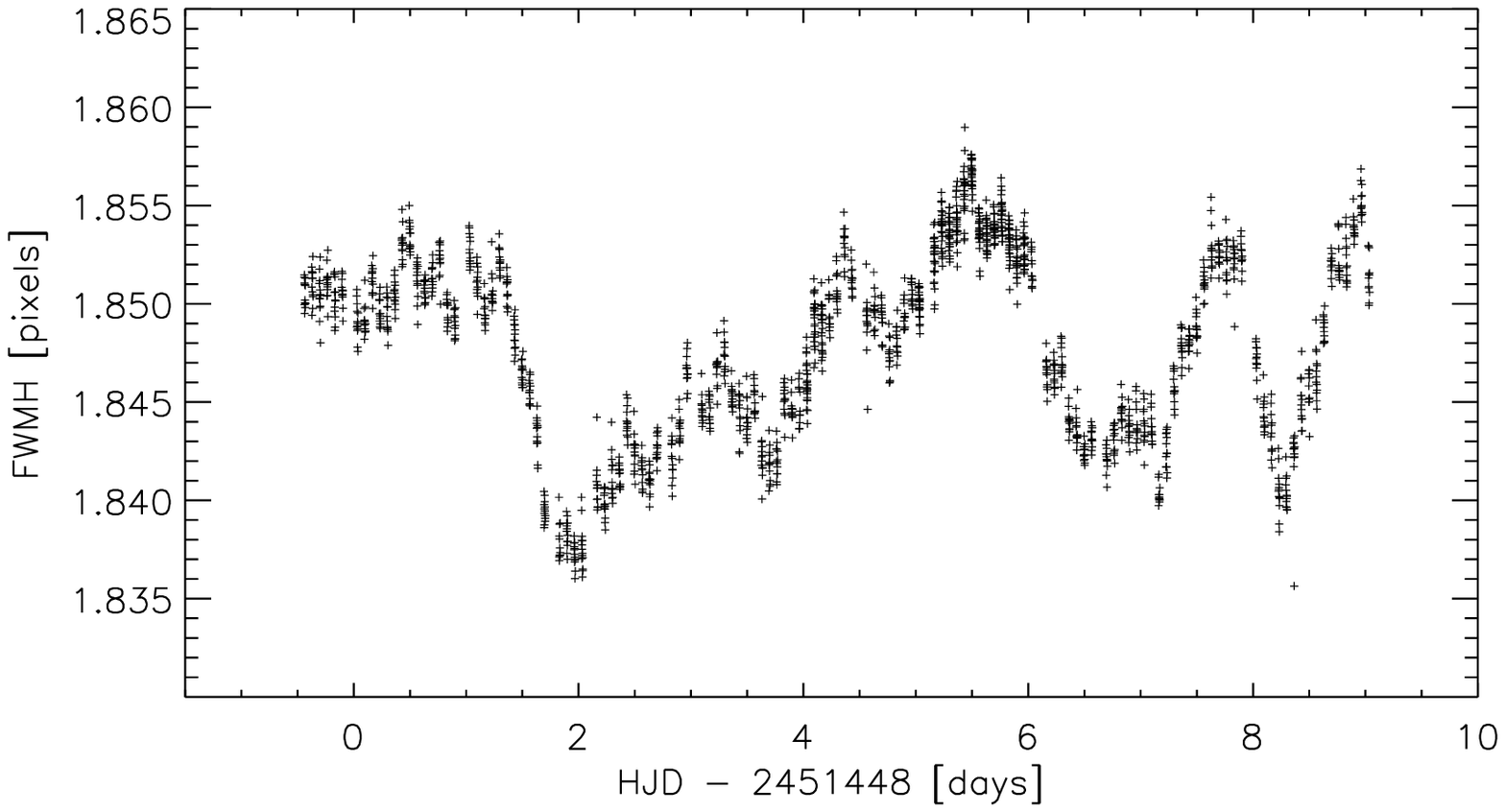} % lc1999_fwhm2.eps}
   \plotone{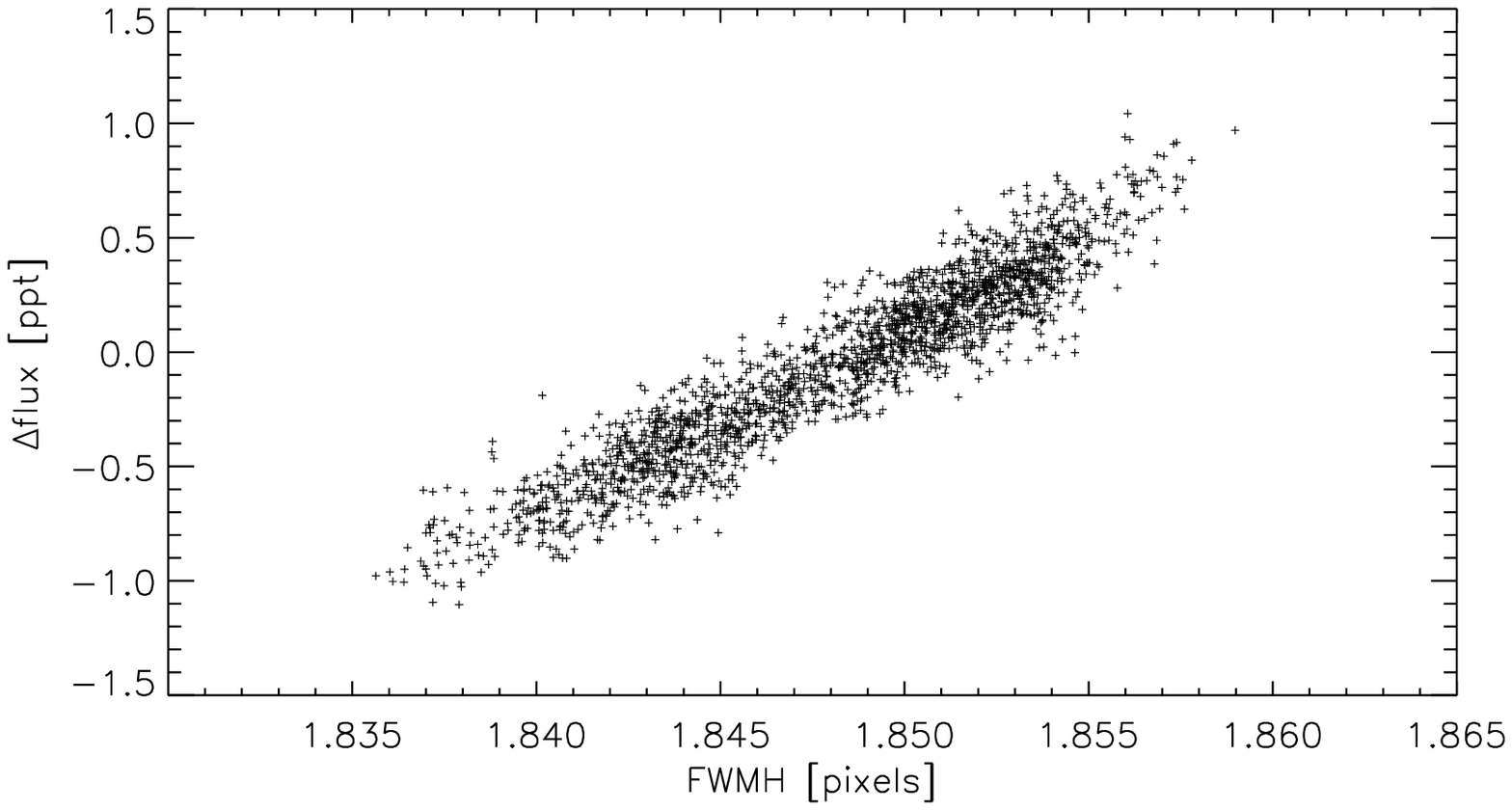} % lc1999_fwhm.eps}
   \caption{The \emph{top} plot shows the change in the 
measured FWHM of the stellar profile versus time for the 
WIRE 1999 light curve. Only data points not affected by
scattered light are plotted and only every 20th data point 
is plotted. In the \emph{bottom} plot the flux change is plotted
versus FWHM when using the same data points as in 
the \emph{top} plot. There is a significant correlation which 
was taken out by a spline fit to the data points.
    \label{fig2}}
   \end{figure}

The {star tracker} on WIRE has a 
52\,mm aperture with a 512$\times$512 CCD. 
A window of 8$\times$8 pixels centered on the star image is integrated for
0.5 seconds and read out.
The FWHM of the stellar image
is around 1.8 pixels and thus is well sampled.
The CCD image is not flat-fielded, so variations of
sensitivity with position are expected. Fortunately, 
the pointing of the satellite is stable to about one 
hundredth of a pixel over periods of several days.
Note that one pixel corresponds to about one arc minute.

WIRE observed Procyon in September/October 1999 and September 2000 for 
9.5 and 7.9 days, respectively. Each of the two data sets consists of 
a few hundred thousand CCD images.
Each WIRE observation consist of five 0.1 second integrations 
which are summed on-orbit by the firmware. %% over which I we no control). 
Since the onboard electronics is 16-bit, when these five
frames are summed for a bright target such as Procyon, 
saturation occurs, but this is
only saturation in the sense that the data registers 
overflow and wrap when reaching $2^{16}=65536$~ADU.
Procyon is so bright that three of the central pixels
reach the digital 16-bit maximum. Above this level
the pixel value will start counting from zero. Thus we added 
$2^{16}$ to the central three pixels to account for this. 

%% Old version from when HB thought the saturation was 'real' saturation:
%% We note that around 50\% of the total light of the star is 
%% contained in these pixels and some degree of
%% non-linearity is expected in the signal. 
%% However, we expect the magnitude of this effect on our results 
%% to be negligible.

We have developed a new pipeline for the data reduction which
differs significantly from our previous work on WIRE data 
\citep{buzasi00, schou, retter} and so we will describe it
in some detail. For the background estimate we used 
12 pixels (3 pixels in each corner of the CCD window).
We fitted a Gaussian to each image to find the approximate position 
and FWHM of the image. For robustness, the background, 
position and FWHM used for each image were the means of 20 images taken before and 
after that image (while requiring that no image was separated
in time by more than 30 seconds). Thus, we typically
averaged these parameters over $41\times0.5=20.5$ seconds. 
We then performed aperture 
photometry with nine increasing sizes, allowing us to select the one 
with the lowest intrinsic noise. The aperture sizes were scaled to the 
measured FWHM to include the same amount of light, independent of FWHM.
However, the changes in FWHM for the Procyon observations were
below 2\% and so the actual number of pixels being summed did not vary. 
For the reduction of the Procyon data we summed 22 pixels. 

%%% MOST time series downloaded from: http://www.astro.ubc.ca/MOST/
%%% BackGr. problem: since part of the light from the star is still present 

For faster processing of the light curves, we binned data within every 
15.5 seconds (\ie\ 31 data points) which resulted in light curves from
WIRE 1999 and 2000 series with 22\,013 and 14\,179 data points. In Fig.~\ref{fig1} we
show the light curve from 1999 for four orbits. In the \topp\ panel the
data are plotted versus time and in the \bott\ panel they
are phased with the orbital frequency of $15.003\pm0.001$~c/day. The
gray points are data affected by scattered Earth light. This is due to
the fact that we underestimated the contribution of the background light 
and thus overestimated the brightness of the star. 
We have tried to remove the scattered light by fitting a spline to the
phased light curve in Fig.~\ref{fig1} but the 
overall noise level is worse than if we just
discard these data. Thus, the data affected by
scattered light have been removed
and the number of data points was then reduced to 
10\,610 and 8\,397 for the WIRE 1999 and 2000 light curves.

In the case of the WIRE 1999 light curve we found that a group 
of data points taken around heliocentric Julian date 
$2451051.3\pm0.4$~days were offset by a significant 
amount, \ie\ $+3$~ppt. We have found no correlation with any of our
reduction parameters, \eg\ FWHM, background level or orbital phase.
This group of points contains 13\% of the total data set and so,
instead of rejecting them, we artificially offset 
these points to have the same median level as the rest of the dataset.
We presume that this effect could be caused by one of the bits
``sticking'' in the ADC.

In Fig.~\ref{fig2} we show how the 1999 photometry correlates with the measured
FWHM of the stellar profile. The \topp\ panel shows the FWHM
changing slowly with time and the \bott\ panel shows flux
versus FWHM\@. There is a significant correlation in the
\bott\ panel, which we removed using a spline fit. 
A similar decorrelation procedure was carried out for 
the WIRE 2000 data.

   \begin{figure}
   \epsscale{.80}
   \plotone{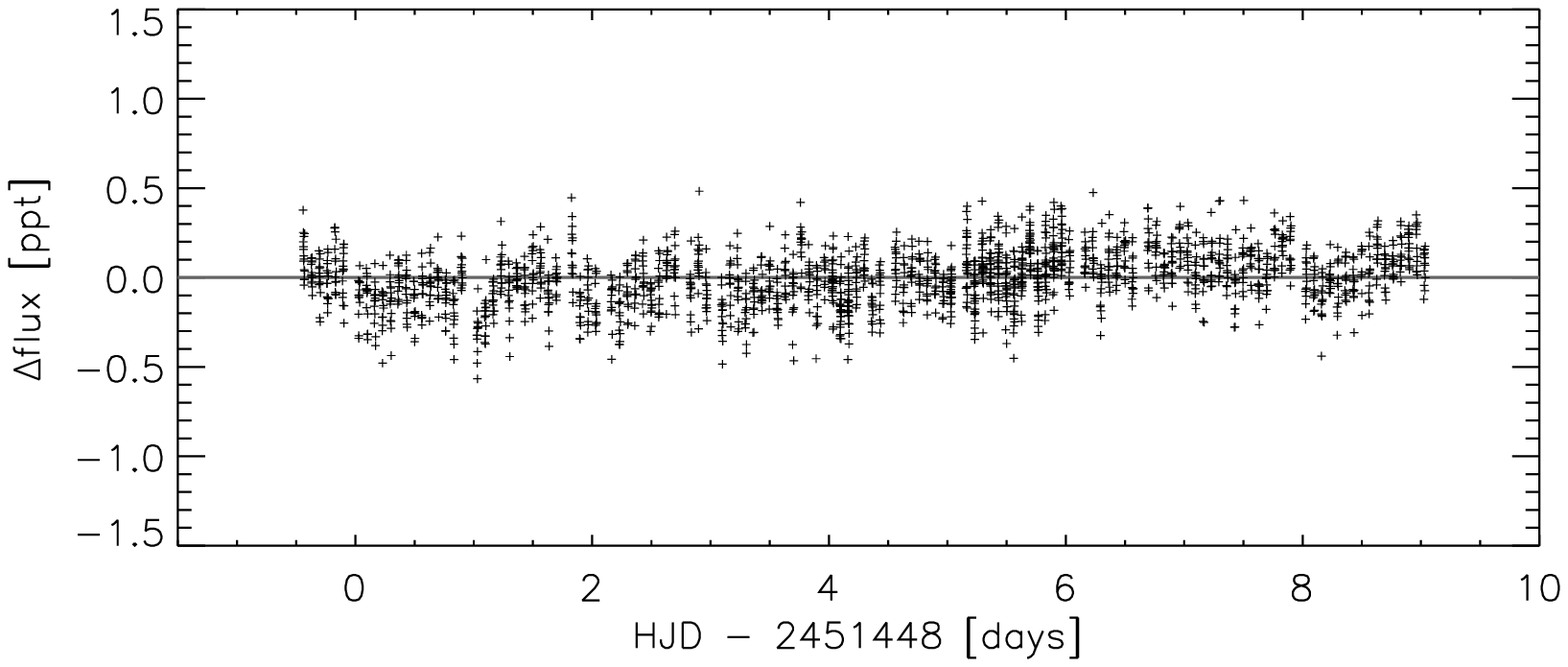} % lc1999.eps}
   \plotone{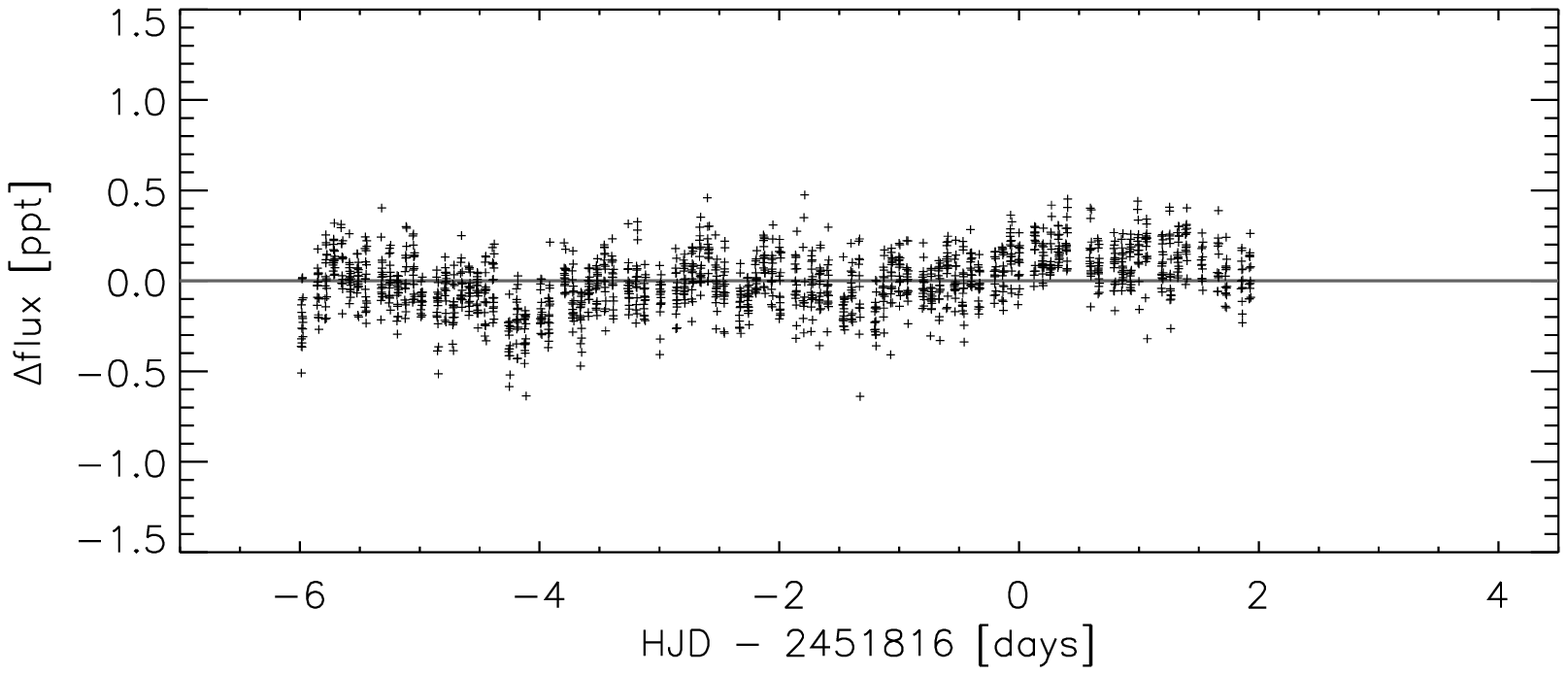} % lc2000.eps}
   \plotone{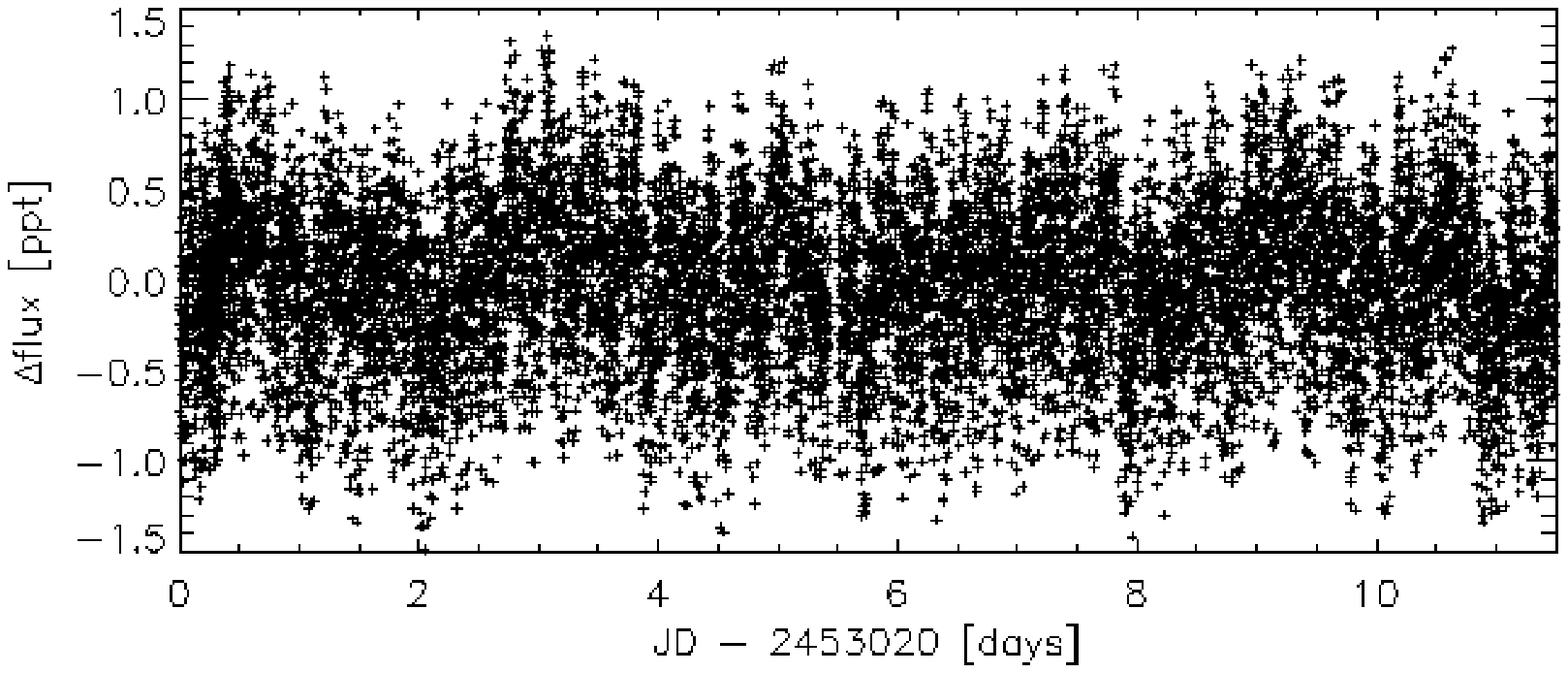} % lc_most2003.eps}
   \caption{The final light curves of Procyon as seen from WIRE in 
September 1999 (\emph{top}) and September 2000 (\emph{bottom}). Data
affected by scatted light has been removed and the correlation 
with FWHM has been removed.  
In each panel, only every fifth data point is plotted.
    \label{fig4}}
   \end{figure}

% ======================================================================
% ======================================================================
% ======================================================================
\subsection{Comparison of the WIRE and MOST time series}
% ======================================================================
% ======================================================================
% ======================================================================

In Fig.~\ref{fig4} we show the final WIRE light curves of Procyon as observed
in 1999 (\topp\ panel) and 2000 (\midd\ panel). For comparison
we also show part of the light curve observed by 
MOST in 2004\footnote{The light curve was obtained at CADC through the MOST website: 
http://www.astro.ubc.ca/MOST/}.
A slowly varying trend seen in the MOST light curve was subtracted. The same
trend was also seen in a comparison star as shown by \citet{matthews04}.
During the first 20 days the MOST time series has a data point every 14.7 seconds 
which is part of the data shown in Fig.~\ref{fig4}. This is very close to the
cadence for the binned WIRE light curves which has one data point each 15.5 seconds.
Note, however that the duty cycle is 18\% in the WIRE datasets compared to 
99\% for the MOST dataset. 
In Fig.~\ref{fig4} only every fifth data point is plotted.

The values of \emph{rms} noise in the 1999 and 2000 WIRE light curves are
135 and 129~ppm, and the point-to-point scatters (which measure
high-frequency noise) are 105 and 101~ppm.  
The photon-noise limit for
WIRE is 88~ppm, which is about 15--20\% lower than the observed
point-to-point scatter.  The extra noise could be due to intrinsic
variations in Procyon or instrumental effects.  For MOST, the
point-to-point scatter is 370~ppm and the photon-noise limit is 111 ppm,
based on the mean flux level of $1.34\times10^7$ ADU \citep{matthews04} and
a gain of 6.1 $e^-$/ADU \citep{walker03}.  Thus, the photon noise in the
MOST data is a factor of 3.3 lower than the observed noise level.
%
%% The MOST light curve has 203\,835 data points and the theoretical limit on
%% the noise level in the amplitude spectrum is 0.33~ppm.
%
Note that the results of \citet{matthews04} were based on a preliminary
data reduction and a more refined reduction gives slightly better results
(J.\ Matthews \& R.\ Kuschnig, private communication).  We conclude that the MOST time
series is affected by a noise source other than pure photon statistics,
which was also suggested by \citet{bedding05} based on the distribution of
peak heights in the amplitude spectrum.

We measured the white noise levels in the 1999 and 2000 WIRE amplitude spectra
at frequencies between 8 and 10\milhz\ and found 1.8 and 1.9~ppm.  Thanks to
the much higher number of data points, the MOST amplitude spectrum has a noise
level of 1.6 ppm in the same frequency range. 

% ----------------------------------------------------------------------------------
% Program: wire_fig5.pro
   \begin{figure}
   \epsscale{.80}
   \plotone{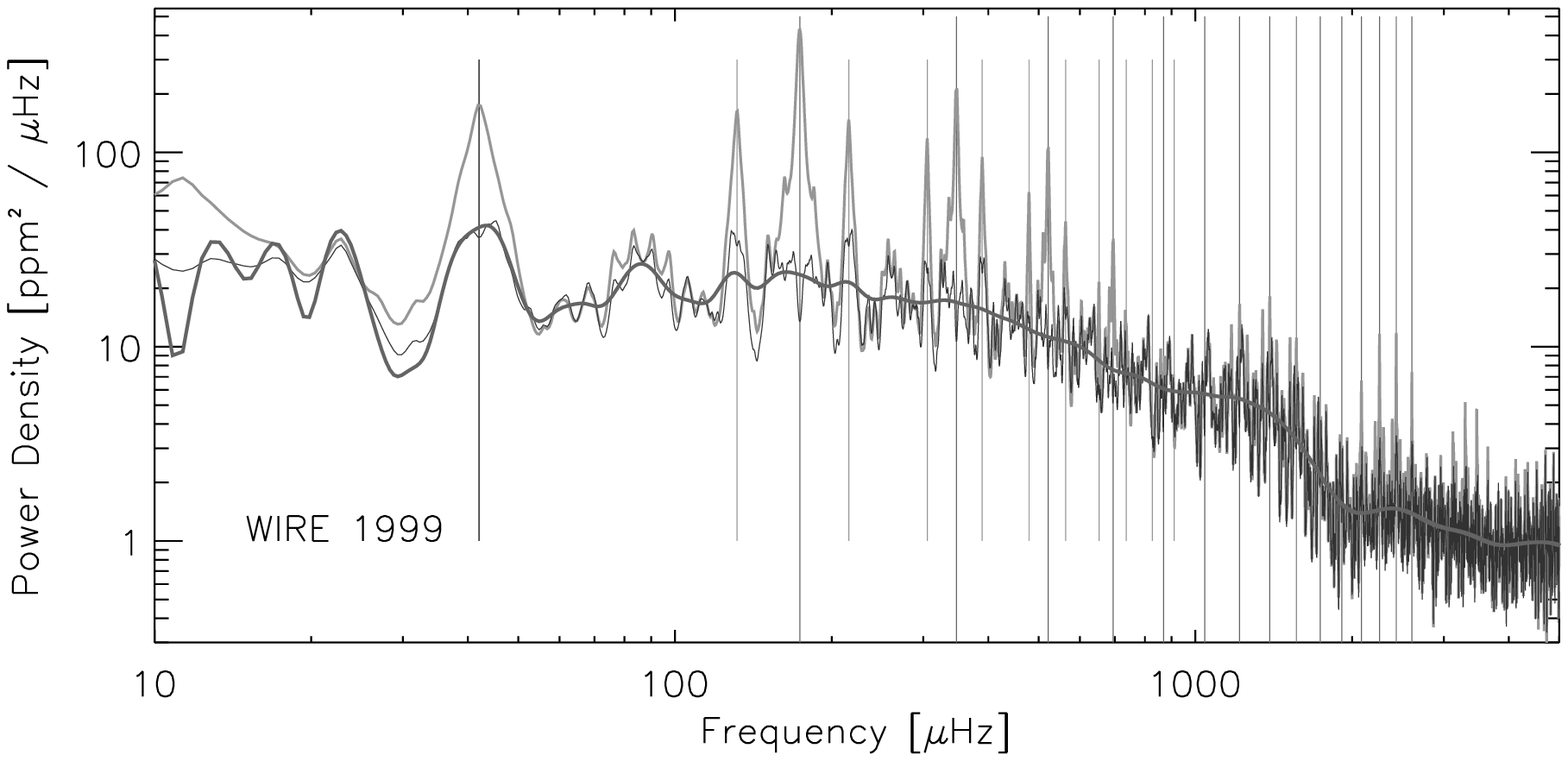} % Procyon_Fig20_1999all.ps} % wire_fig20 with option: addyear='1999all'
   \plotone{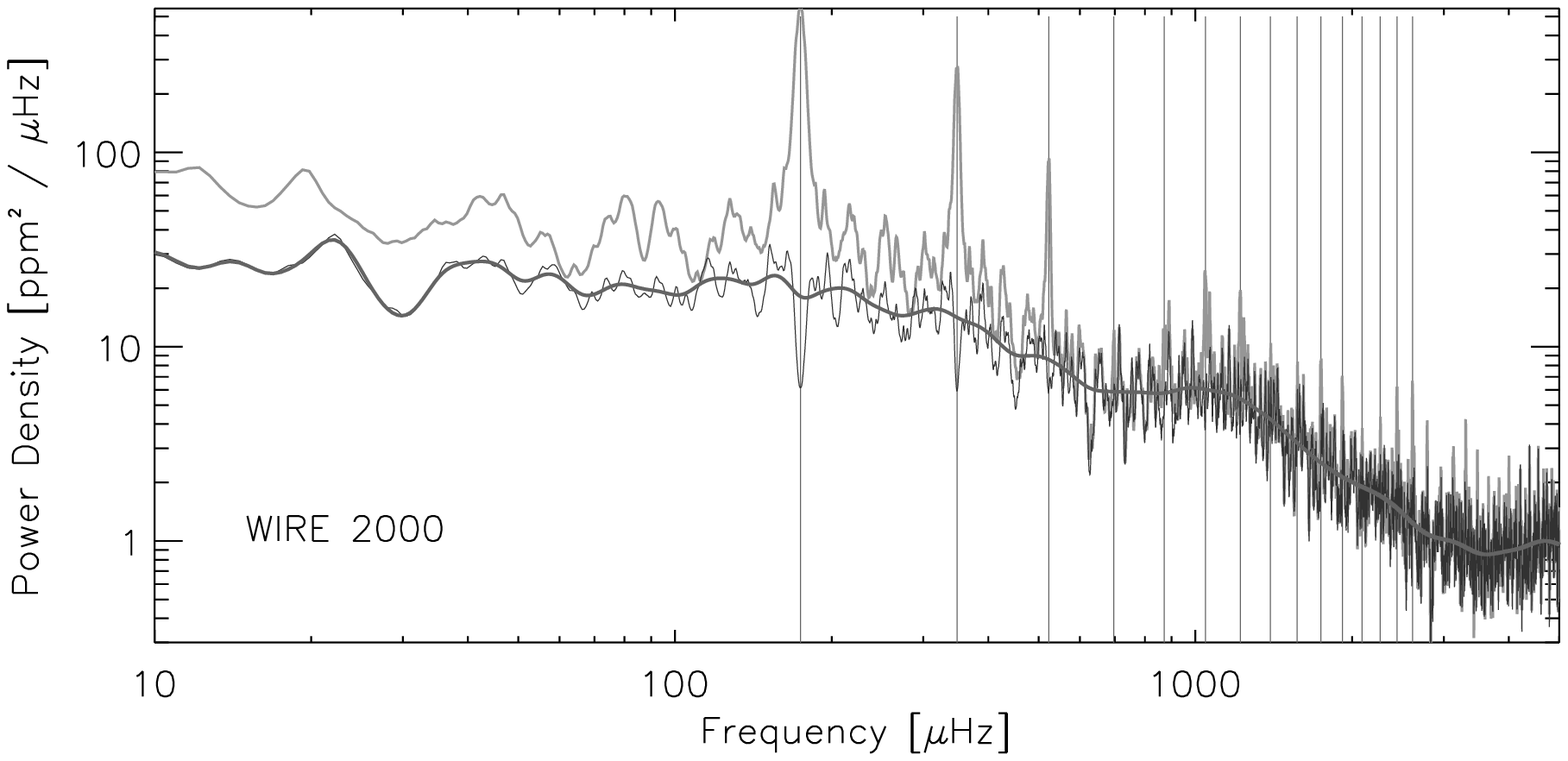} % Procyon_Fig20_2000all.ps}
   \caption{Power density spectrum of the WIRE 1999 (\topp) and WIRE 2000
(\bott) light curves. The gray curves are the original PDS and the black
curve is the PDS after subtracting low frequency components. The smoothed
version of the cleaned spectrum is also shown. The vertical lines mark
integer multiples of the orbital frequency. For the WIRE 1999 light curve
the combination frequencies of the peak at $\sim43$\milhz\ are also marked.
    \label{fig:density}}
   \end{figure}
% ----------------------------------------------------------------------------------

% wire_fig5_plot.pro with option: addyear = '1999 + 2000'

   \begin{figure}
   \epsscale{.80}
   \plotone{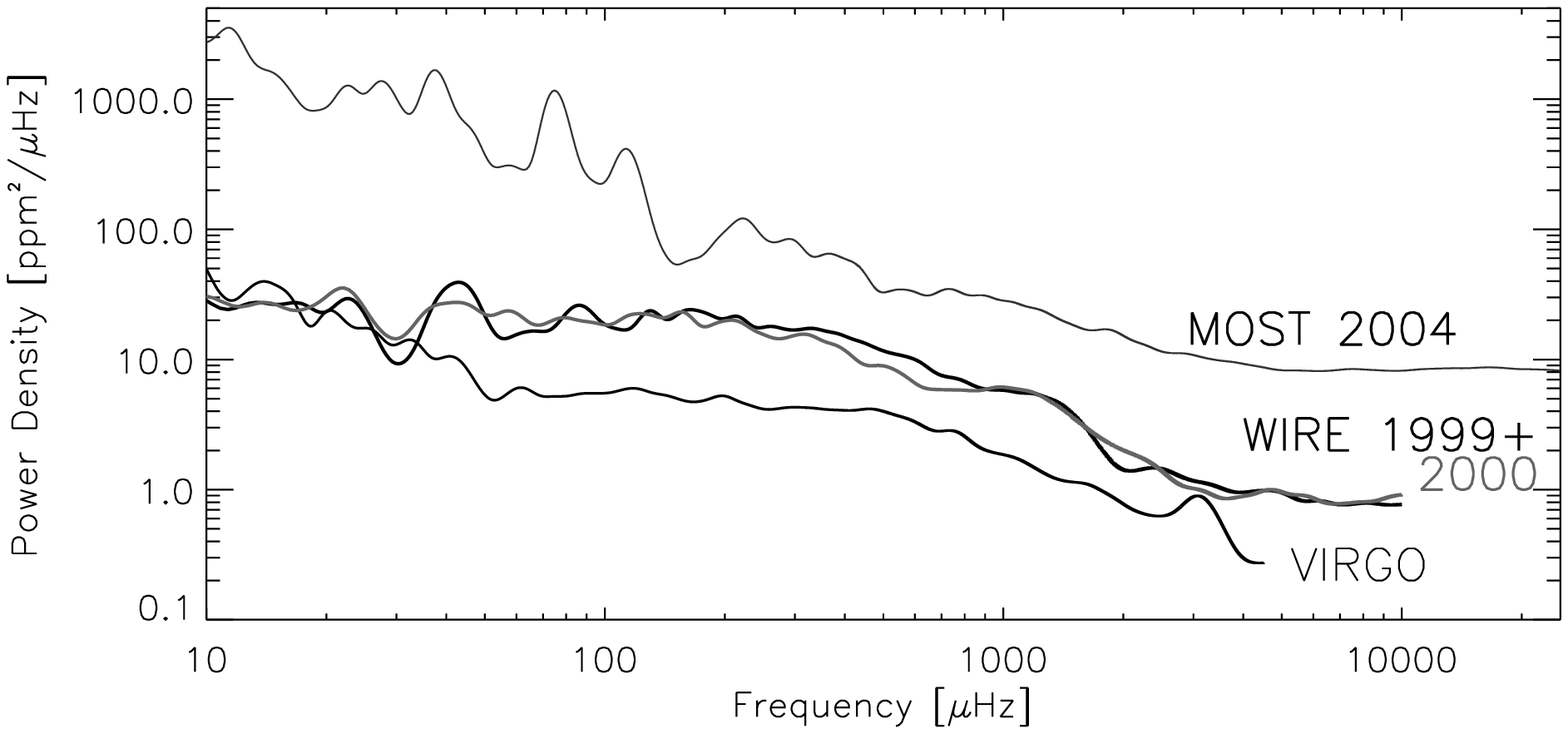} % Procyon_Fig5_1999+2000.ps}
   \caption{Smoothed PDS based on the MOST 2004, WIRE
1999 and 2000 data sets.  The PDS based on VIRGO solar data from the green
channel is shown for comparison.
    \label{fig6}}
   \end{figure}

% ===========================================================================
\section{Power density analysis\label{sec:pow}}
% ===========================================================================

%   In this Section we will examine the properties of the two Procyon
%   light curves in the frequency domain. First we will compare our 
%   results with the results from the MOST\footnote{Micro-variability and
%   Oscillations of STars; \citet{walker03}} satellite \citep{matthews04} 
%   before comparing the WIRE results with simulations of the 
%   granulation and p-modes in Section~\ref{sec:sim}.
%   
%   % ===========================================================================
%   \subsection{Comparison with the MOST spectrum} 
%   % ===========================================================================

To measure noise levels as a function of frequency in the different time
series, we calculated power density spectra (PDS).  Power density measures
the power per frequency resolution element, which automatically makes it
independent of the length (and sampling function) of the time series.  We
calculated the frequency resolution by measuring the area under the
spectral window, which is the power
spectrum of a single sinusoidal oscillation with an amplitude of 1~ppm that is
sampled using the same times and weights as the observed series.  These areas,
which we used to convert from power to power density, were 7.29 and 7.02 \mhz\ for
the WIRE 1999 and 2000 light curves, respectively.

The WIRE light curves contain a number of
significant outliers and to suppress their influence on the amplitude
spectra we have computed weights for all points.  Each point was a weight
based on the scatter relative to the neighboring four data points 
(with the additional constraint that those data points must be within the same
WIRE orbit).  We then smoothed these weights using a running mean of width
150 data points.

The gray curve in each panel of Fig.~\ref{fig:density} shows the PDS from
WIRE for 1999 and 2000.  There are a few peaks below 15 \mhz\
(corresponding to periods longer than one day) which we do not attribute to
the star.  This low-frequency power generates aliases close to the harmonics of
the orbital frequency of the WIRE satellite, around $n\times174$~\mhz,
which are marked by vertical lines.  In the case of the WIRE 1999 spectrum,
we also see a power excess around 43~\mhz\ that is not seen in the WIRE
2000 data.  This excess also appears as aliases on either side of each
orbital harmonic, as marked by additional vertical lines in the upper panel
of Fig.~\ref{fig:density}.

We removed the low-frequency components from each light curve by the
standard method of repeatedly subtracting the sinusoid corresponding to the
strongest peak in the power spectrum.  We removed three and four components from
the WIRE 1999 and 2000 light curves, respectively, and so obtained the high-pass filtered
PDS shown in black in Fig.~\ref{fig:density}.  We can see that the excess
power at the harmonics of the orbital frequency is greatly reduced.  In
each panel we have also over-plotted a smoothed version of the high-pass
filtered PDS, and these are also shown in Fig.~\ref{fig6}.

It can be seen in Fig.~\ref{fig:density} that the high-pass filtered WIRE 1999  
PDS still has some excess power at the orbital harmonic frequencies that
is not seen in the WIRE 2000 PDS. This may explain the slightly 
higher level seen in the PDS of WIRE 1999 compared to WIRE 2000 
in the 300--800\mhz\ range in Fig.~\ref{fig6}.

In Fig.~\ref{fig6} we also compare the WIRE PDS with that from the MOST data.
We calculated the frequency resolution of the MOST time series, as
described above, to be 0.412~\mhz, which is close to the theoretical value
$10^6/(32 \cdot 0.99 \cdot 86400)$\mhz\ $= 0.365$\mhz\ for a 32 day time
series with a 99\% duty cycle.  From Fig.~\ref{fig6} we find 
that the noise level in the range 200--600\mhz\ in the WIRE data 
is a factor of 4 lower in power (2 in amplitude) compared to the MOST data.

In Fig.~\ref{fig6} we also show the power density spectrum 
based on VIRGO data (green channel) of the Sun \citep{frohlich}.
The granulation signal in the Sun around 100--300\mhz\ is about 
a factor of two lower in power compared to the WIRE observations of Procyon.
The bump seen at $\sim3$~\mhz\ is power from the solar p-modes.

When comparing observations, we must keep in mind that they were
taken over different wavelength ranges.
The MOST satellite has a broadband filter covering 350--700~nm,
while the VIRGO green channel has a narrow band at 500~nm. 
The filter response of the WIRE star tracker is not known, 
but calibrations of the count levels observed in about 50 stars of 
different spectral type indicate that the
filter response is more sensitive in the blue than in the red.
We find the observed count levels observed with WIRE 
agree for both early and late type stars if the filter is 
Johnson~$B$. Thus, the passbands of the MOST, VIRGO and WIRE 
spacecraft are roughly comparable.

% \citet{matthews04} claimed that the high noise level they observed
% could be due to the granulation. As shown by \citet{bedding05} the
% granulation noise should thus be four times higher than solar 
% in amplitude. However, the results in Fig.~\ref{fig6} shows that the
% granulation noise must be much lower in Procyon by a factor of 
% $\simeq1.9$ in amplitude.

% Factor to divided power spec by: dummy_fac = (1e6/(32. * 0.99 * 86400.))
% = 0.365 microHz

% ===========================================================================
\section{Simulations of Procyon\label{sec:sim}}
% ===========================================================================

% Program: wire_procyon_simmode.pro:
   \begin{figure}
   \epsscale{.80}
   \plotone{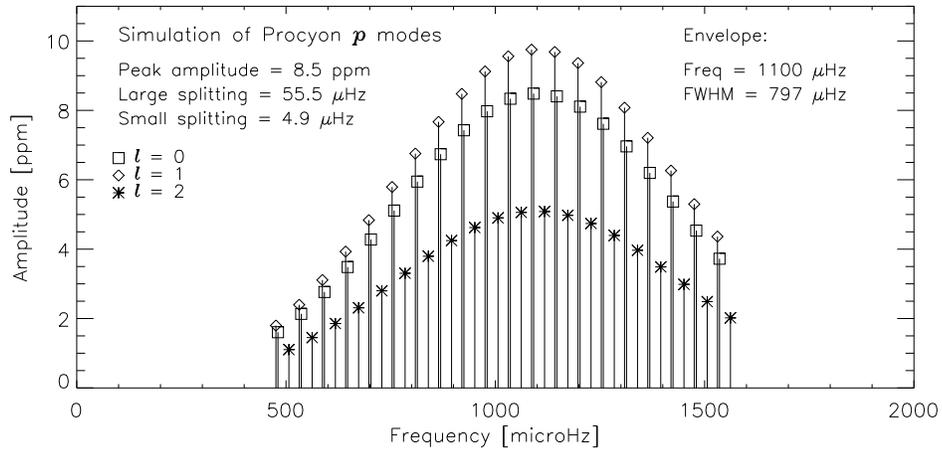} % Procyon_Fig_Pmodes.ps}
   \caption{Input frequencies and amplitudes for the
simulation of p-modes in Procyon.
    \label{fig:modes}}
   \end{figure}

% Programs: wire_fig7, 8, 9 and wire_fig10.pro
   \begin{figure}
   \epsscale{.45}
   \plotone{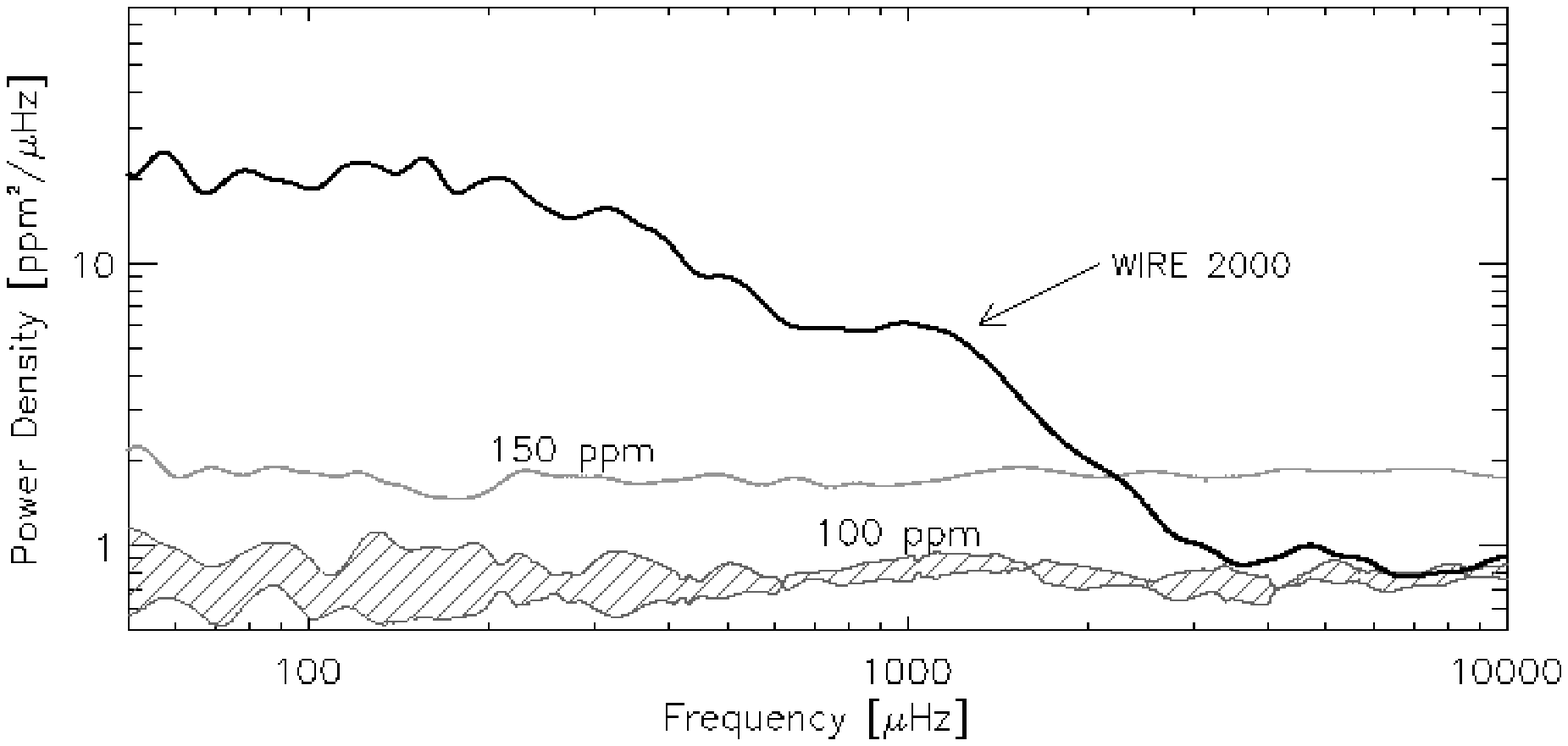} % Procyon_Fig7_2000.ps}
   \plotone{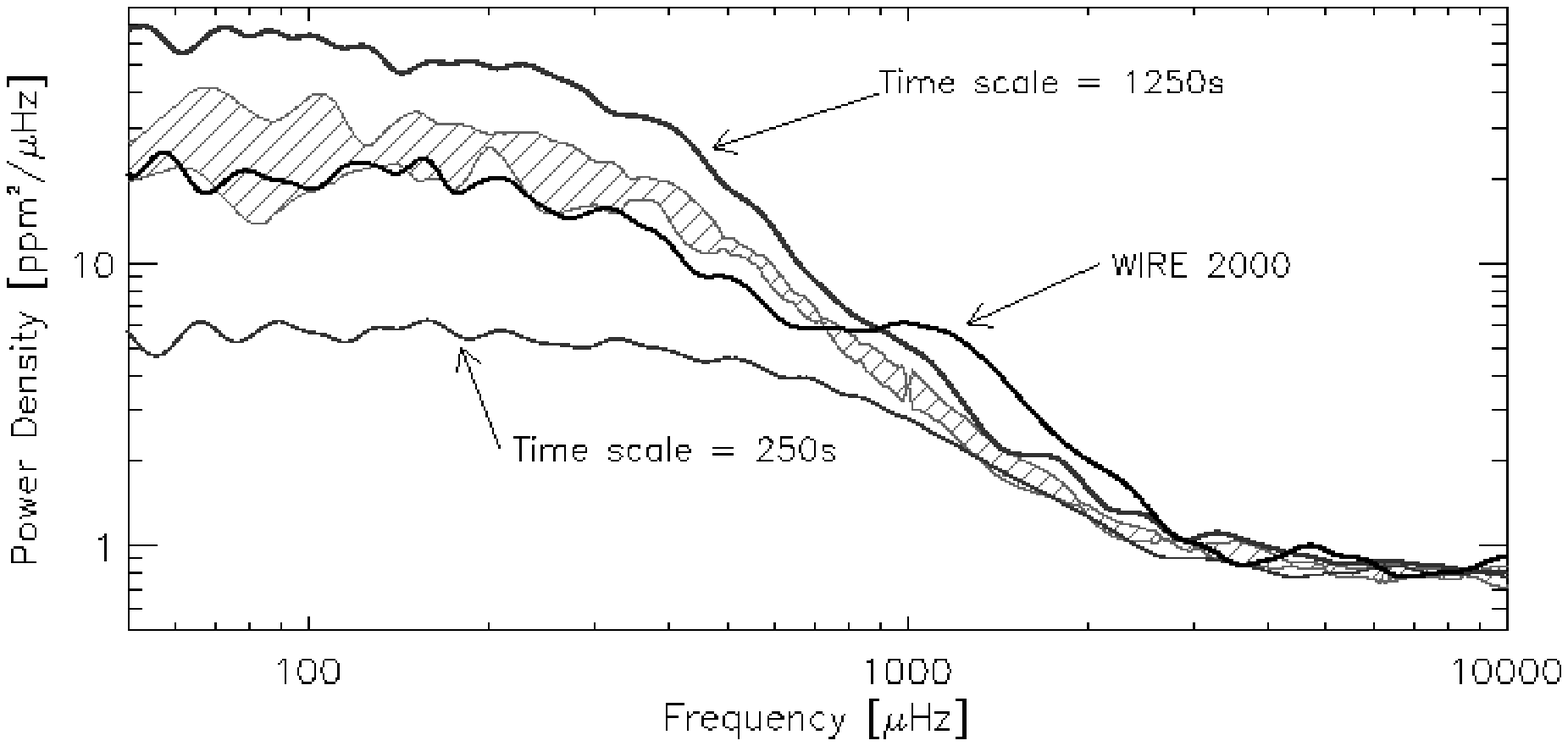} % Procyon_Fig8_2000.ps}
   \plotone{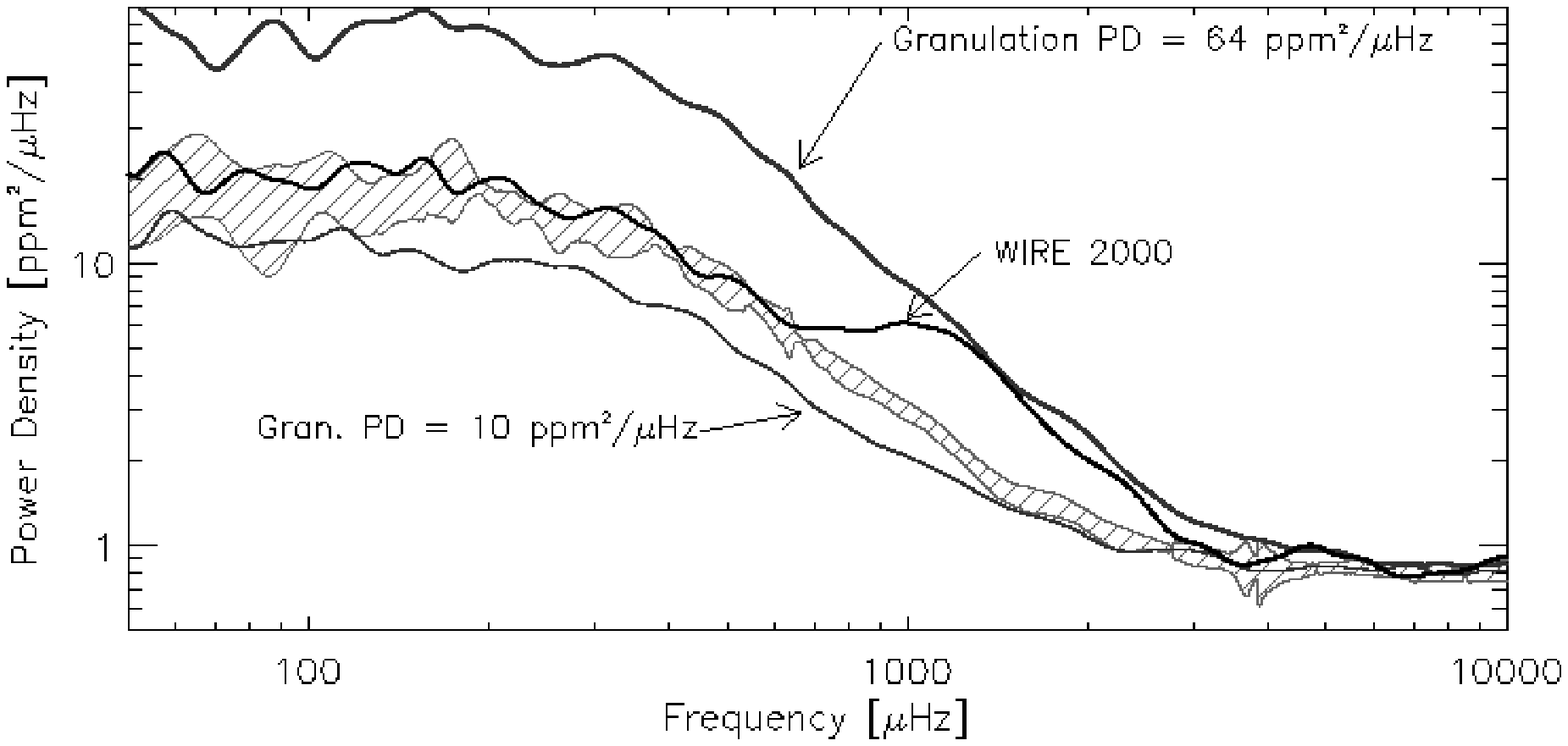} % Procyon_Fig9_2000.ps}
   \plotone{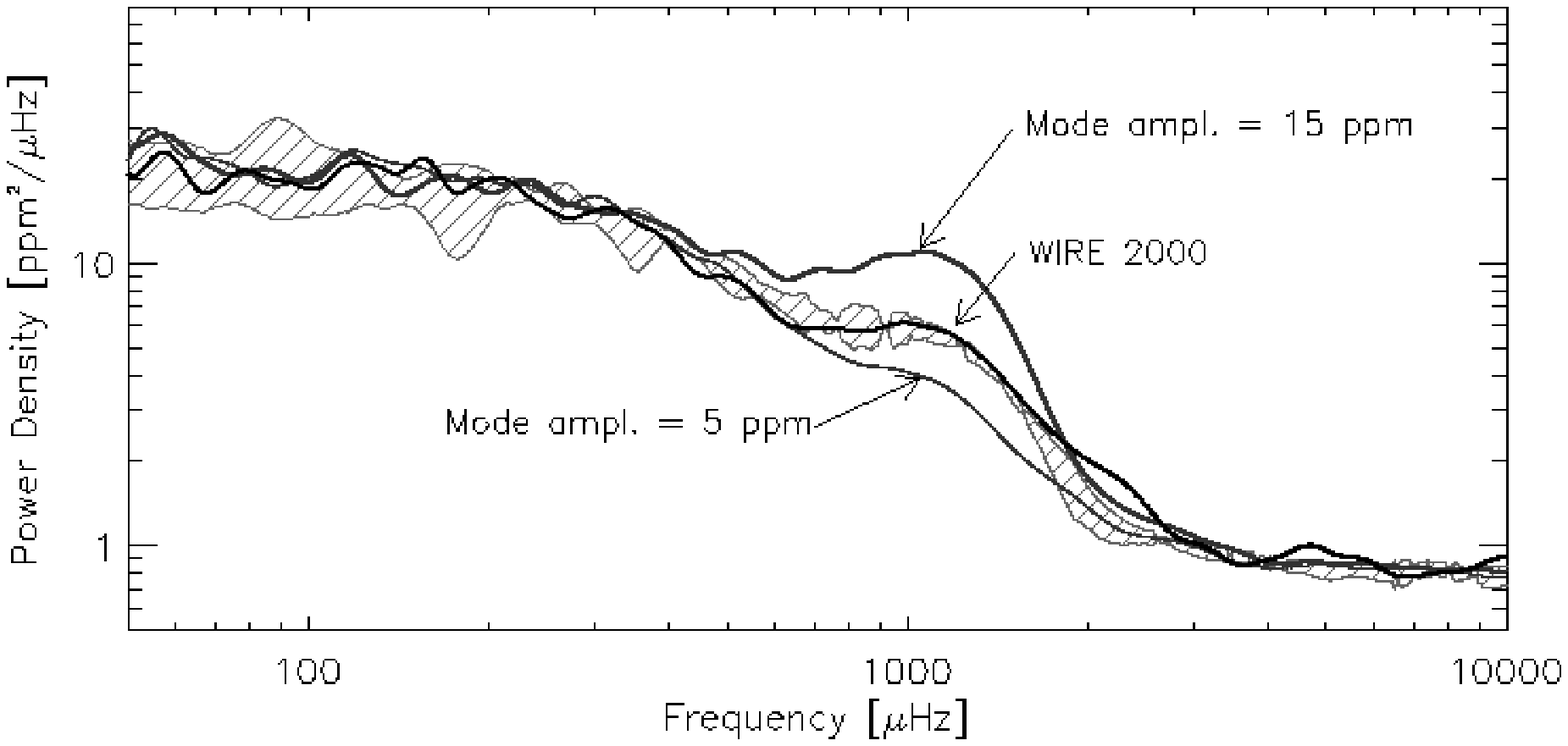} % Procyon_Fig10_2000.ps}
   \caption{The four panels show the
power density spectrum of the WIRE 2000 along with
different simulations. Each simulation is the mean
of five simulations with different seed numbers. 
The hatched regions show the 1-$\sigma$ variation
for selected simulations. In the \topp\ \lee\ panel 
simulations for two different white noise levels
are shown.
In the \topp\ \rii\ panel the simulations have 
timescales of the granulation of 250, 750, and 1250\,s. 
In the \bott\ \lee\ panel the timescale of the granulation
is 750\,s but the granulation power densities (PDs) are 10, 18, and 64 ppm$^2$/\mhz.
%%% ampG ==  25, 34, and 65~ppm.
In the \bott\ \rii\ panel the granulation timescale
and granulation PDs are 750\,s and 18~ppm$^2$/\mhz,
%%% 34 ppm,
while the amplitude of the p-modes are 5, 10, 15~ppm. 
    \label{fig:sim}}
   \end{figure}

We have made simulations of the granulation and
p-modes in Procyon based on a 
software package developed to provide 
time-series simulations for the Danish MONS/R\o mer
satellite mission as well as for the 
ESA Eddington mission. We refer 
to \citet{stello04} for details of 
the simulations.

The simulation software has several parameters that can
be modified. 
Firstly, we can set the white noise level in the simulations, which
is determined by the noise level at high frequencies (above 8\milhz). 
For the granulation noise, we can change the timescale
and granulation amplitude, while the p-modes are described by
peak amplitude, lifetime and frequencies. 
The granulation signal seen in the PDS will depend on
the input amplitude but also the time scale of the granulation.
Thus, we use the measured granulation power density 
at low frequencies (20--120\mhz) from this point on.
To include the p-modes in the simulation, we adopted a 
large separation of 55.5\mhz\ and a small separation of 4.9\mhz\
\citep[e.g.][]{martic04}. 
We only included modes of degree $l=0,1,$ and $2$ and we have 
modified the amplitudes by a Gaussian envelope 
to emulate the signal seen in other sun-like 
stars.  % (see \eg\ \citep{kjeldsen2004}).  
As an example, we show the distribution of the 
amplitude of the modes for peak amplitudes 
for the $l=1$ modes of 8.5~ppm in Fig.~\ref{fig:modes}. 

\subsection{Estimating the parameter range}

To get an idea of the parameters that are needed to fit the observed PDS,
and of the uncertainties in these parameters, we have made several
simulations.  In Fig.~\ref{fig:sim} we show results for the WIRE 2000 time
series.  In each panel, the observed spectrum is shown as a solid black
line, and we adjusted the parameters in the following four steps.

The \topp\ \lee\ panel shows in gray two simulated PDS with white noise
levels of 100 and 150~ppm, respectively.  In this and all other panels, the
hatched region shows the 1-$\sigma$ variation, which was found from five
simulations that differed only in the seed for the random-number
generator. For clarity, this 1-$\sigma$ variation is only shown for one of
the cases.  Based on this panel, we have fixed the white noise at 100~ppm
for the remaining three panels.

In the \topp\ \rii\ panel we have now
added granulation with three different timescales: 250, 750 and 1250
seconds.  These simulations allow us to rule out timescales lower than
250\,s, based on the frequency at which the slope changes.
For the next two panels, we have fixed the granulation timescale at 750\,s.

In the \bott\ \lee\ panel, having fixed the white noise and the granulation
timescale, we now show the PDS for three different granulation levels:
10, 18, and 64 ppm$^2$/\mhz. Note that the granulation
power density is measured at low frequencies, \ie\ below 120 \mhz.

Finally, in the \bott\ \rii\ panel we retain the previous parameters and
fix the granulation level at 18~ppm$^2$/\mhz.  %% at ampG = 34~ppm.  
We now add p-mode oscillations
with three different peak amplitudes: 5, 10, and 15~ppm.  We adopted a mode
lifetime of 1 day but note that, given the way the PDS are smoothed, the
results are not affected by changing this value.  

Our simulations are able to fit the observed power density spectrum quite
well.  Based on Fig.~\ref{fig:sim} we can conclude that the white noise
level is about $100\pm5$~ppm, the granulation timescale must be larger than
250\,s and probably less than 1250\,s, the granulation PD must lie
around $18\pm5$~ppm$^2$/\mhz\    %%% $34\pm5$~ppm, print, 0.01515 * 2. * 34. * 5.
while the peak amplitude of the p-modes must
certainly be less than 15 ppm.

% =============================================================
% =============================================================
% =============================================================
\subsection{Constraining the simulation parameters\label{sec:constrain}}
% =============================================================
% =============================================================
% =============================================================

% Program: wire_fig12.pro 
   \begin{figure}
   \epsscale{.45}
     \plotone{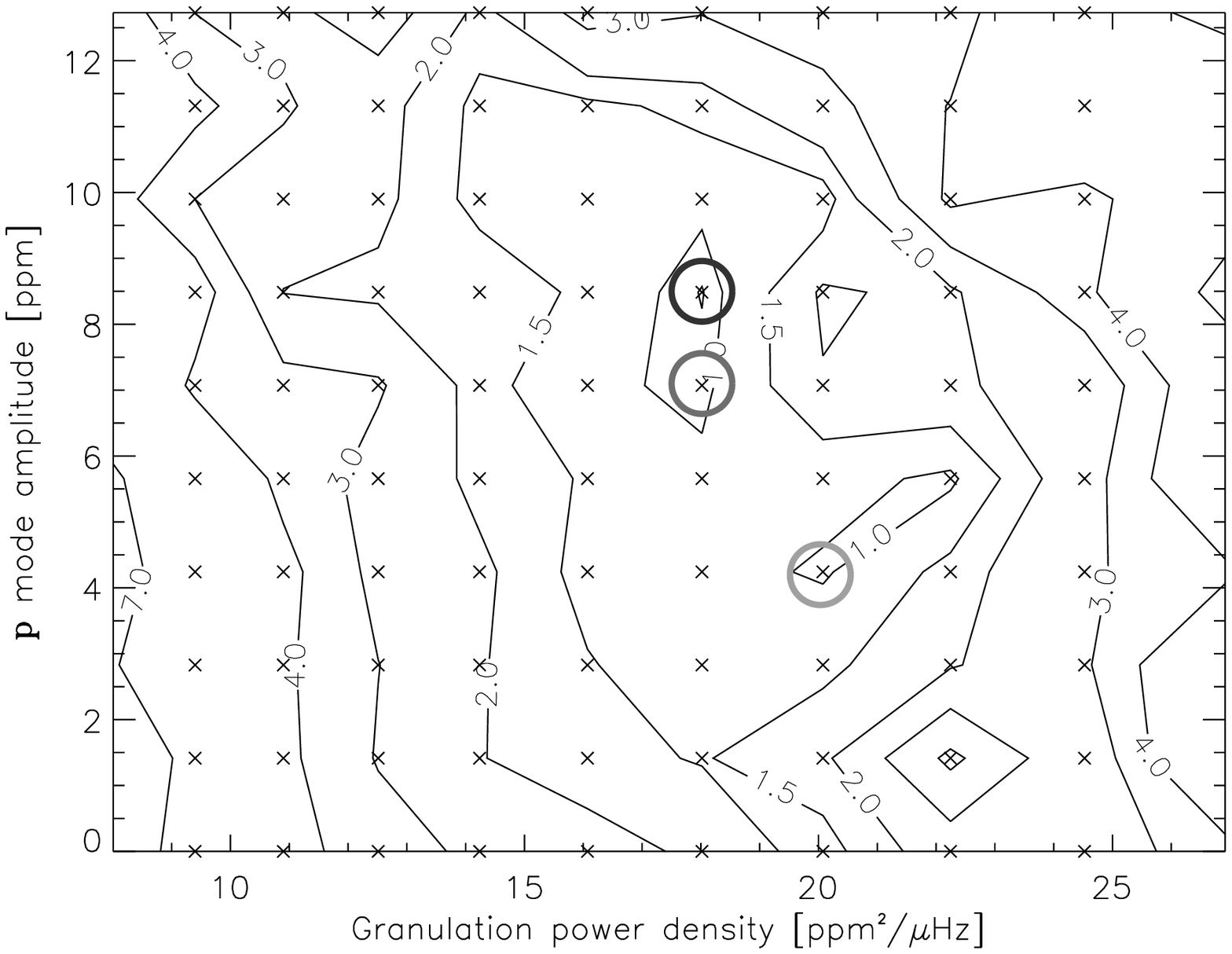}  % Procyon_Fig12_2000_500s_1.0d_new.ps}
     \plotone{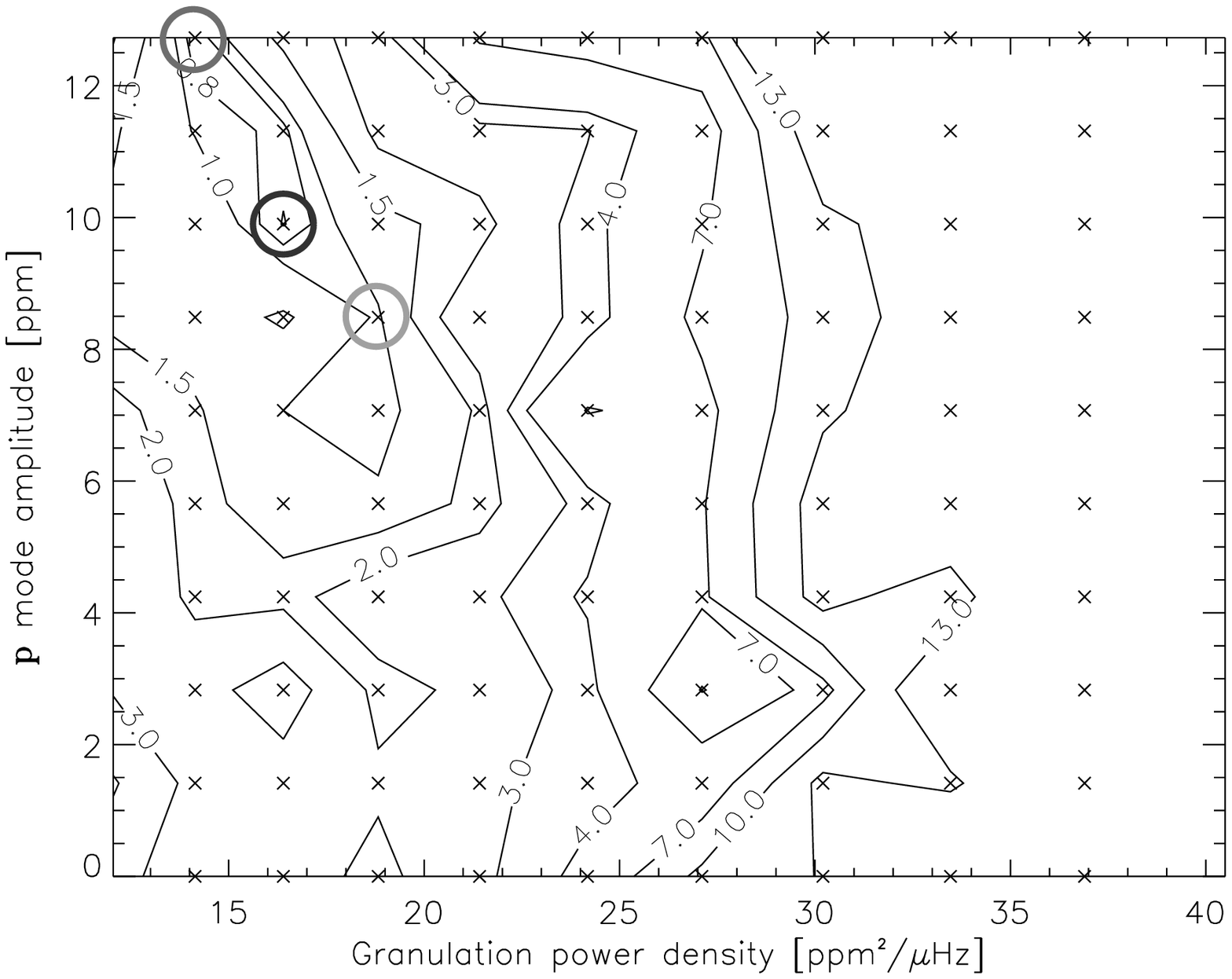}  % Procyon_Fig12_2000_750s_1.0d_new.ps}
   \caption{The two panels show 
   contours of the $\chi^2$ for the comparison of the
   WIRE 2000 and simulation power spectra with
   different values for the granulation PD
   and p-mode amplitude. The granulation time
   scales are 500 (\lee\ panel) and 750 seconds.
    \label{fig:con}}
   \end{figure}

% Program: wire_fig15.pro (best models are plotted)
   \begin{figure}
   \epsscale{.70}
   \plotone{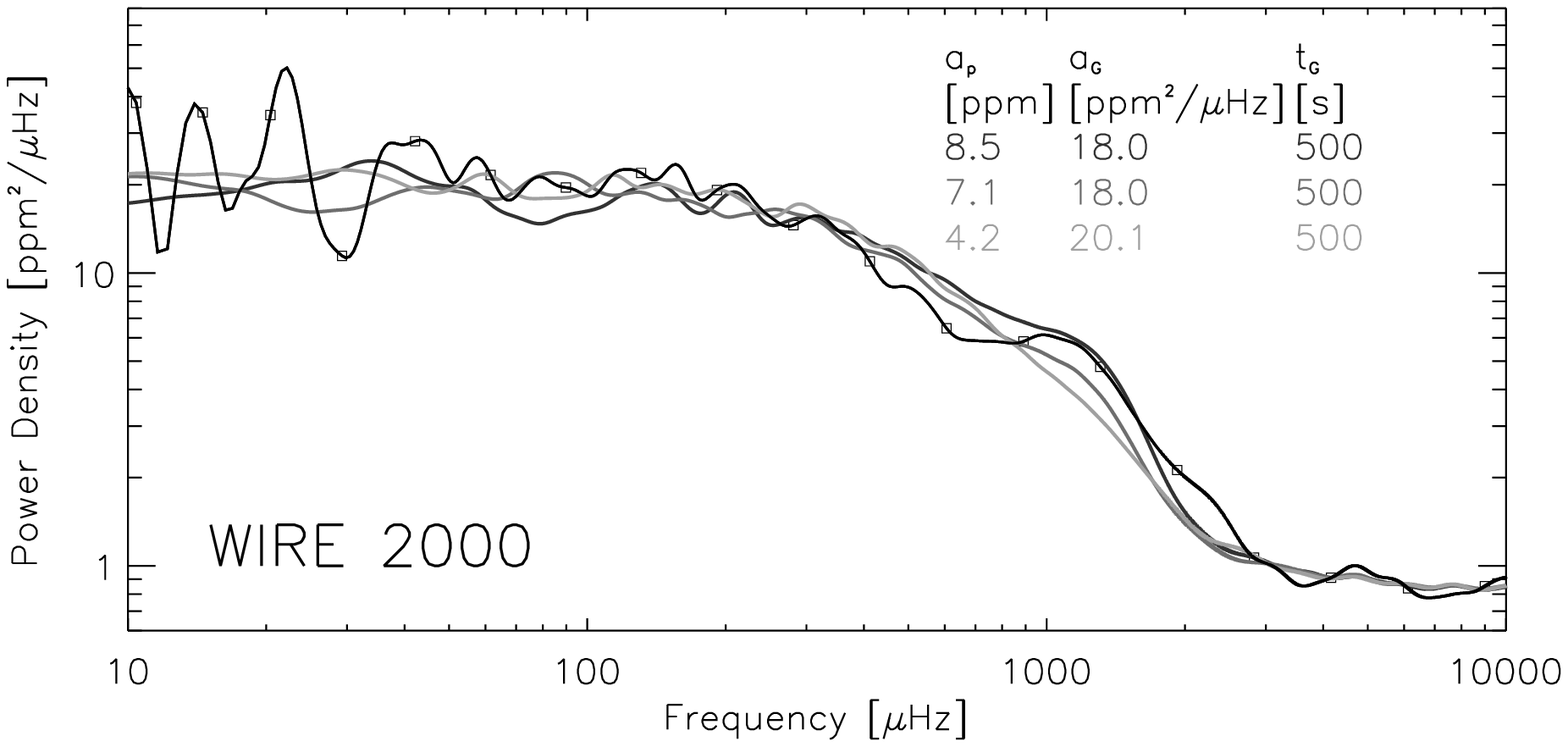}  % Procyon_Fig15_2000_tg500_sev.ps}
   \plotone{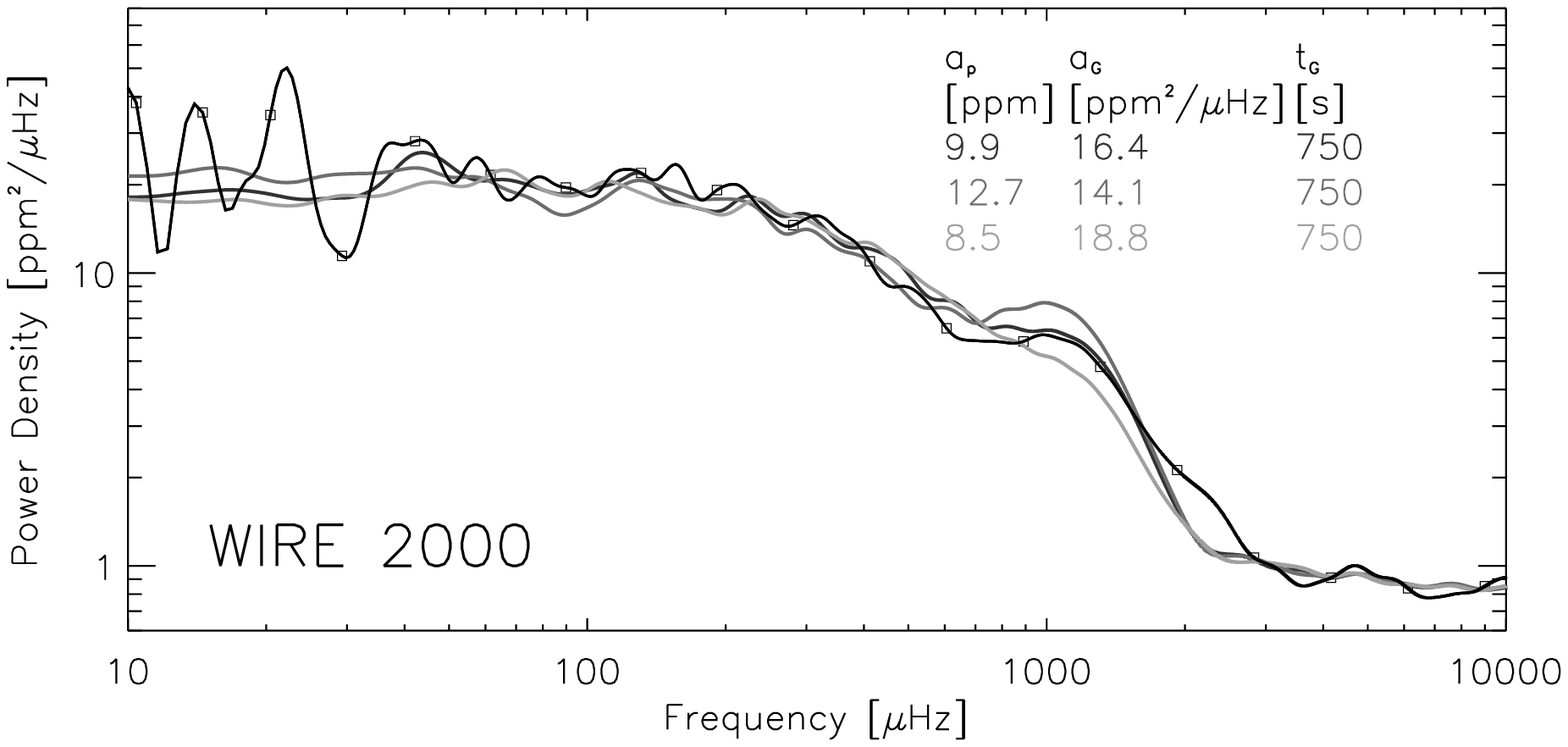}  % Procyon_Fig15_2000_tg750_sev.ps}
\caption{The power density spectra of the WIRE 2000 light curve
compared to models with granulation timescales of 500 (\upp\ panel)
and 750 seconds. The WIRE 2000 spectrum is marked by open box symbols.
The different shades of gray correspond to the 
filled circles on the contour plot in Fig.~\ref{fig:con}. The parameters
of the models are given in the upper right corner of each panel.
    \label{fig:goodsim}}
   \end{figure}

% Program: wire_fig15.pro (best models are plotted)
   \begin{figure}
   \epsscale{.70}
   \plotone{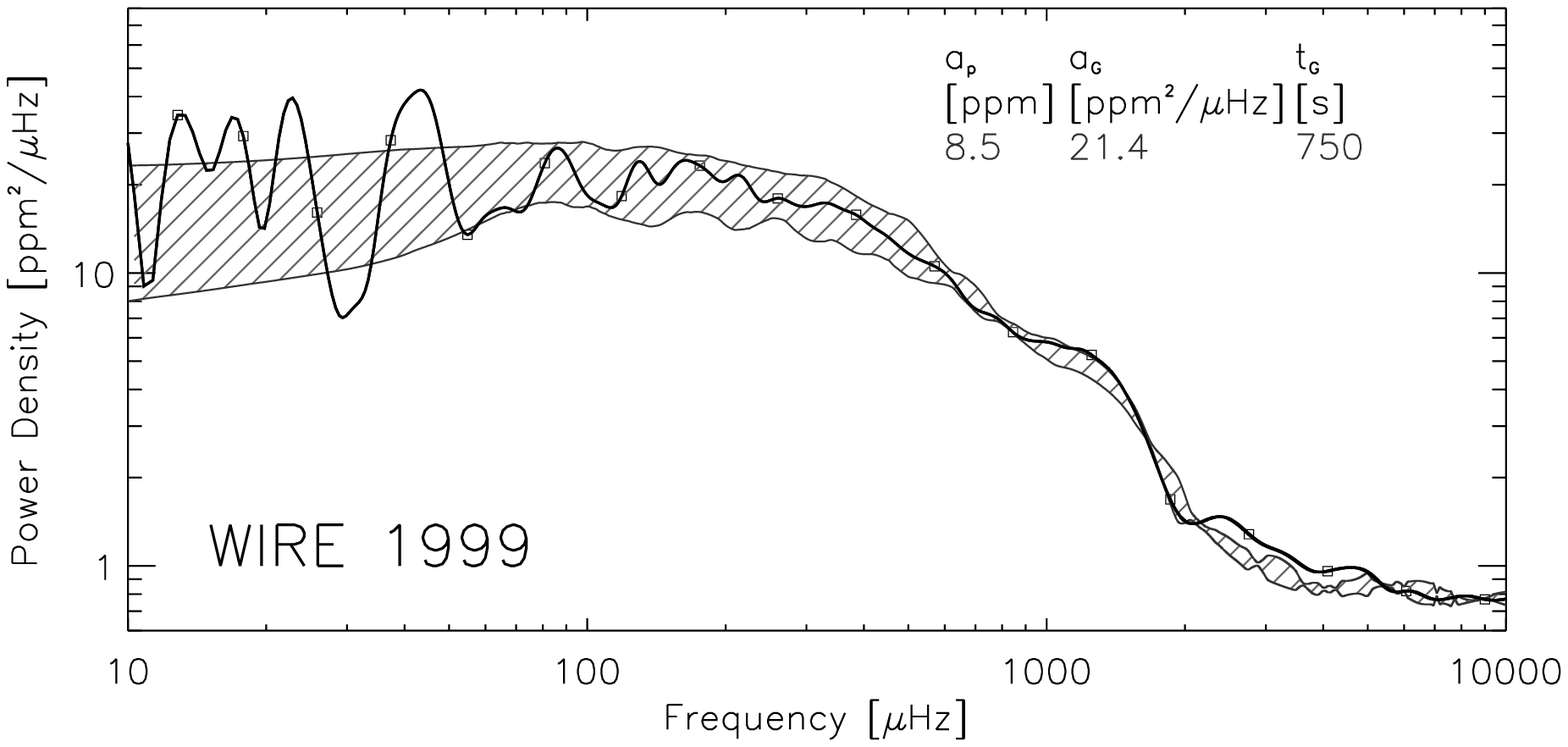}  % Procyon_Fig15_1999_tg750_best.ps}
   \plotone{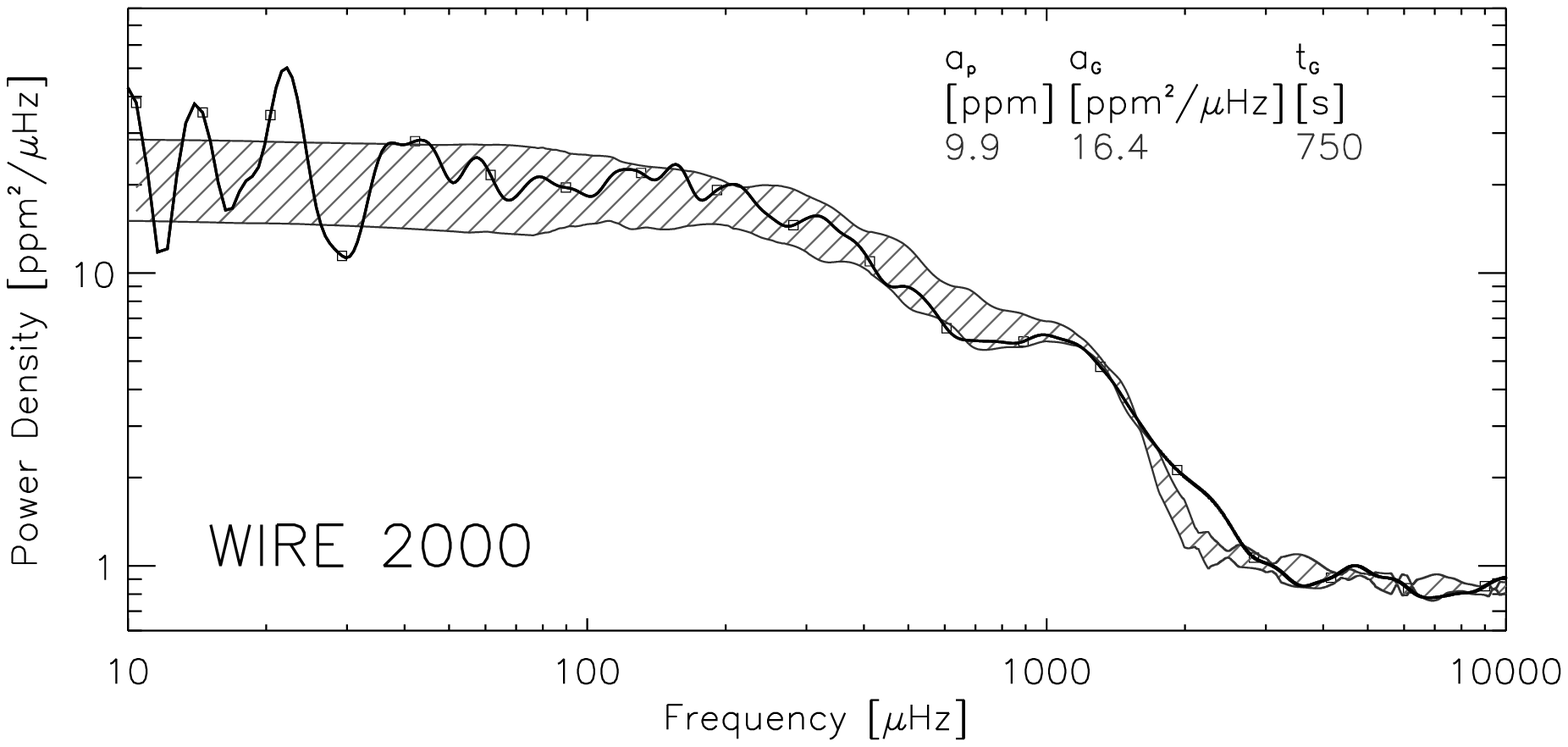}  % Procyon_Fig15_2000_tg750_best.ps}
\caption{The power density spectra of the WIRE 1999 (\upp\ panel)
and WIRE 2000 light curves and the models with the best fit. 
The WIRE spectra are marked by open box symbols and 
the 1-$\sigma$ variation of the simulations is shown 
by the hatched region. 
    \label{fig:bestsim}}
   \end{figure}

\clearpage

% tgran: [500,750,1000.,1250] seconds
% agran: [.6,.65,.7,.75,.8,.85,.9,.95,1.,1.1,1.05]  * 47 ppm
% Solar == 30 ppm, range: 0.04 .. 1.65 times solar
% amode: [4,6,8,10,12,14,16,18] / sqrt(2) ppm ; 2.8 to 12.7 ppm
% lifet: [0.5,1.0,2.0] days

% Step size in granulation amplitude:
% a = [.6,.65,.7,.75,.8,.85,.9,.95,1.,1.1,1.05]  * 47.
% print,a(1:10) - a(0:9)

% Step size in p-mode amplitude
% b = [0,2,4,6,8,10,12,14,16,18] / sqrt(2) & print,b(1:7) - b(0:6)

% Parameter range for grid updated 2nd of Feb. 2005 by HB

%% print, [28.,52.]^2. * 0.01515 
%% Step size changes with ampG:
%% print, ([28.]^2.-[28.+2.4]^2. ) * 0.01515 ; 2.12
%% print, ([52.]^2.-[52.+2.4]^2. ) * 0.01515 ; 3.87

To put further constraints on the granulation PD and oscillation
amplitudes, we have computed a large grid of simulations 
that include the ranges of parameters indicated above. 
We made time series with granulation timescales 
of 500, 750, 1000, and 1250\,s; 
granulation PD in the range 12--41 ppm$^2$/\mhz\ in average 
step size of 3~ppm$^2$/\mhz;
   %%% granulation amplitudes in the range 28--52 ppm in steps of 2.4 ppm;
and oscillation amplitudes between 0 and 13 ppm in steps of 1.5 ppm.
The white noise was fixed at 100 and 105~ppm for 
the WIRE 1999 and 2000 time series, respectively. 
Each time series was calculated for five different
random number seeds to estimate the scatter. 
In total we calculated $5\times440=2200$ 
time series for both the WIRE 1999 and 2000 datasets.

For each time series we calculated the power density spectrum.
In order to select the best models we computed
the $\chi^2$ in the frequency range between 50 and 5000\mhz.
We used logarithmic frequency bins in order to avoid
giving too much weight to the high-frequency part of the spectrum. 

In Fig.~\ref{fig:con} we show results for the simulations of the
WIRE 2000 time series. The two contour plots correspond to
granulation timescales of 500 and 750\,s, respectively.
The contours show the $\chi^2$ value for each set of granulation PD
and oscillation amplitudes.  The crosses mark the
locations of the computed simulations. We have marked three models
with circles of different color that have the smallest values of $\chi^2$, and
the corresponding power density spectra are shown in
Fig.~\ref{fig:goodsim}, where
the simulation parameters are given in each panel. 
We find that the region 400--800~\mhz\ is better fitted for the
granulation timescale of 750\,s. 
This region is located just below the excess power seen around 1\milhz.

In Fig.~\ref{fig:bestsim} we show examples of simulations that
provide good fits to the WIRE 1999 and 2000
power density spectra. The hatched regions correspond to 
the 1-$\sigma$ uncertainty, based on the scatter of the five simulations with
different random number seeds. The agreement between 1999 and 2000 for the
best-fitting parameters is 
not surprising given the similarity of the power density curves
(cf. Fig.~\ref{fig6}). The main difference between the two WIRE PDS
is the region 400--800~\mhz. 
%% As was noted in Sec.~\ref{sec:pow}, the higher level of the WIRE 1999
%% spectrum in this range may due to an insufficient subtraction of 
%% peaks at low frequency. 
As noted in Sec.~\ref{sec:obs}, we found a group
of data points that were offset from the rest in the WIRE 1999 light curve
that may be the source of additional noise.

%%  (to be investigated further by HB!).

%% print, 35.^2. * 0.01515, 0.01515 * 2. * 35. * 4.

We thus conclude that the parameters that
best fit the observed power density curves lie in the range
$750\pm200$, $18\pm4$~ppm$^2$/\mhz,      %%% $35\pm4$~ppm, 
and $8.5\pm2$~ppm for the
granulation timescale, granulation PD and p-mode
amplitude, respectively. 

The granulation time scale in Procyon is about twice that of the
Sun (VIRGO data; \cf\ Fig.~\ref{fig6}),
the granulation PD is three times higher, 
and the peak p-mode amplitude is twice solar. 

%% (the mean level in VIRGO data set around 100--300~\mhz 
%% is $\simeq5.5$~ppm$^2$/\mhz; \cf\ Fig.~\ref{fig6}), 
%% Thus the granulation PD in Procyon is about $1.8\pm0.3$ times the Sun.

% New section: HB on 17th of March 2005
% ===============================================================
% ===============================================================
\subsection{Peak height distribution\label{sec:peak}}
% ===============================================================
% ===============================================================

% The main feature of the 
% power density spectra shown in Fig.~\ref{fig:density} is that
% the power increases towards lower frequencies, 
% \ie\ from 2\,000 to 200~\mhz. Furthermore there is evidence of 
% excess power around 1\,000~\mhz\ which may be due to the p-modes 
% since it is found in the same frequency range as reported by the
% radial velocity campaigns on Procyon (\eg\ \citet{martic04}).
% Our simulations seem to support this interpretation and we put
% an upper limit to the oscillation peak power at $8.5\pm2$~ppm.

\begin{figure}
   \epsscale{.70}
   \plotone{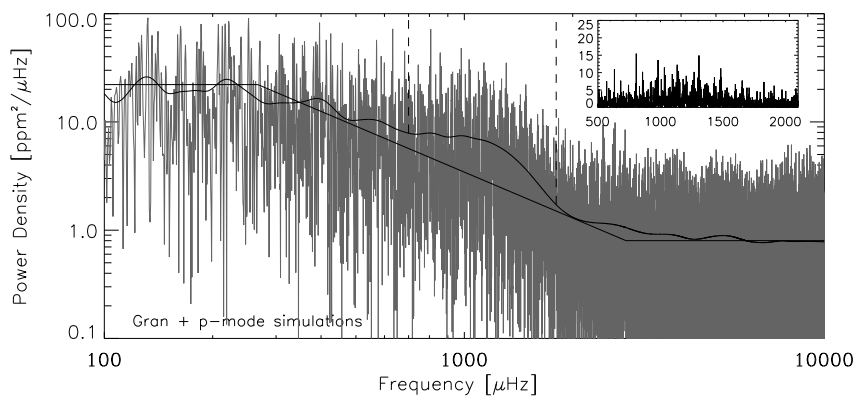}  % PeakNormSim.eps}
\caption{Power density spectrum for one of the simulations of
granulation and p-modes when using the same input times as
for the WIRE 2000 time series. The solid line marks the normalization and the
resulting spectrum is shown in the insert. The dashed lines
mark the frequency region used for the examination of the
peak distribution (\cf\ Fig.~\ref{fig:peakdist}).
    \label{fig:peaknorm}}
   \end{figure}

Following \citet{bedding05}, we have investigated whether the distribution
of peak heights in 
the power density spectrum of the 
WIRE light curves (\cf\ Fig.~\ref{fig:density}) agrees with
the simulations.  We consider
the frequency range 700--1\,800~\mhz, which contains the excess
power that may be due to oscillations. Since the
power increases rapidly 
when going to lower frequencies, we have normalized
the power density spectrum as illustrated in Fig.~\ref{fig:peaknorm}. 
We did this by removing the overall slope by dividing the PDS 
by the straight line shown in the logarithmic plot in Fig.~\ref{fig:peaknorm}.
We have chosen a simulation
of the WIRE 2000 time series with granulation timescale and
granulation PD of 750\,s and 16.4~ppm$^2$/\mhz,    %%% 33 ppm, 
respectively, which gave the best
fit for the PDS 
(\cf\ Fig.~\ref{fig:bestsim}). The p~mode
peak amplitude is 9.9~ppm and the lifetime of the modes is
1 day. The solid line was divided into the spectrum and the
result is shown as the insert in Fig.~\ref{fig:peaknorm}. 
The same was done for the observed power density spectrum. 

% PeakNormObs.eps
% PeakNormSim.eps
% PeakNormWN.eps
% PeakNormWNBump.eps
% PeakNorm10dSIM.eps

% Program: wire_peakdist_norm.pro

% Program: wire_peakdist.pro
   \begin{figure}
   \epsscale{.70}
   \plotone{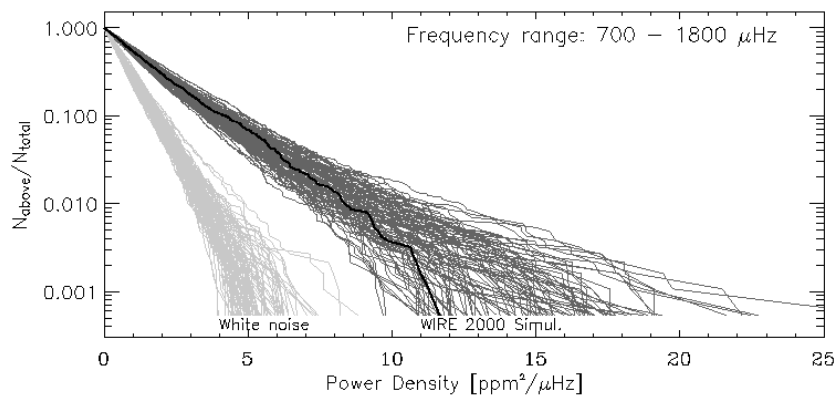}  % Procyon_PeakDist_Neutral.ps}
\caption{The cumulative distribution of peak heights in the
observed WIRE 2000 light curve (single black curve). 
The light gray
lines are simulations of white noise while the dark
gray lines simulations of granulation and 
p~modes. 
   \label{fig:peakdist}}
   \end{figure}

In Fig.~\ref{fig:peakdist} we show the cumulative distribution of
peak heights after this normalization, \ie\ the fraction of peaks 
that are above a given level. 
The cut-off at around $N_{\rm above}/N_{\rm total}\simeq0.0005$ corresponds
to the inverse of the number of data points in the spectrum 
in the frequency range 700--1\,800~\mhz, \ie. we have $N_{\rm total} = 1\,865$.
The light gray lines are results for 100 simulations of pure white noise ,
while the dark gray lines are based on 100 simulations of granulation
and p-modes with the same parameters used for Fig.~\ref{fig:peaknorm}. 
We conclude that the actual WIRE data from 2000 contains an 
excess of strong peaks which is consistent with p-modes. 
The WIRE 1999 data set gives very similar results.

% New section: HB on 17th of March 2005
% ===============================================================
% ===============================================================
\subsection{Search for a Comb-like Structure\label{sec:comb}}
% ===============================================================
% ===============================================================

The p-modes detected in the Sun and other solar-like stars show
a distinct distribution of peaks, often referred to as the comb structure.
The separation between the dominant peaks is twice the large separation and
can be determined from the autocorrelation of the power spectrum. We applied
this method to the WIRE data but did not find evidence for the large
separation. We applied the same analysis to the 
simulations with the parameters found in 
Sect.~\ref{sec:constrain}, also with a negative result. We repeated
the analysis for a simulation with the same duration and time sampling
but with a 100\% duty cycle and did indeed recover the input for the
large separation. However, the S/N was also improved significantly 
since there was about five times more data points in this simulation.
This emphasizes that a high duty cycle is important
when future asteroseismology space missions will observe solar-like stars.

% ===========================================================================
\section{Conclusion\label{sec:con}}

We have analyzed light curves of Procyon from the WIRE satellite 
from two epochs in 1999 and 2000. When comparing the power 
density spectrum (PDS) of
Procyon and the Sun we find a higher noise level by a factor of
$1.8\pm0.3$ at frequencies in the range 100--300~\mhz, which is expected since
Procyon is hotter and more luminous than the Sun.

We find evidence for excess power around 1\mhz\ which is consistent with
the p-modes reported from spectroscopic 
campaigns \citep{brown91, martic04, eggenberger04, claudi05}.
We have compared the observed PDS of the observations with
a grid of simulations with different input granulation timescale,
granulation PD and p-mode amplitude. We have thus constrained the 
granulation timescale, granulation PD and p-mode
amplitude to be $750\pm200$, $18\pm4$~ppm$^2$/\mhz, %%%% $35\pm4$~ppm, 
and $8.5\pm2$~ppm, respectively. The upper limit we put on the
p-mode peak amplitude is consistent with measured radial velocity 
amplitudes which lie in the range 50--70\,cm/s \citep{martic04}.
We note that the determined peak amplitude of the p-modes 
does not depend on the assumed life time (1~day)
but will depend slightly on our assumed distribution of peaks 
(\cf\ Fig.~\ref{fig:modes}). In particular this is true 
for the assumed width of the the Gaussian envelope used for 
modifying the input peak heights.

We have compared the distribution of peak heights in the PDS of 
the observations and in several simulations of granulation
and p-modes. We find a good agreement for the peak distribution which
supports our interpretation of the excess power around 1\mhz\ as being due
to p-modes. 
However, we find no evidence for the comb-like structure
expected for solar-like oscillations. 

Our results do not agree with the much higher noise level 
(a factor two in amplitude) found from observations with the 
MOST satellite \citep{matthews04}. The analysis by \citet{bedding05} of 
the MOST power spectrum showed that the the distribution of peak heights 
was not consistent with a pure noise source and \citet{bedding05} concluded 
that an additional error source could be present. 
This is confirmed by the present comparison with observations from the WIRE satellite.

% We finally note that Procyon has been observed again by WIRE 
% in the spring of 2005, and our team is currently analyzing this dataset.

% ===========================================================================
\acknowledgments
% ===========================================================================

HB and DLB acknowledge support from NASA (NAG5-9318) and
from the US Air~Force Academy. HB and HK are grateful for the
support from the Danish Research Council 
(Forskningsr\aa d for Natur og Univers) and TRB thanks 
the Australian Research Council.
We are grateful to the MOST team for making the Procyon 
data available through the Canadian Astronomy Data Centre
(operated by the Herzberg Institute of Astrophysics, National
Research Council of Canada). We also thank Dennis Stello for
useful discussions and, in particular, his ideas for the investigation
of the peak-height distribution in power spectra.

% ===========================================================================


%% copy editing.  Individual instruments can be provided in parentheses,
%% after the keyword, but they will not be verified.

% Facilities: \facility{Nickel}, \facility{HST(STIS)}, \facility{CXO(ASIS)}.



\clearpage

\end{document}